\documentclass[a4paper,fleqn]{cas-dc}

\usepackage[numbers]{natbib}
\usepackage{amsmath,amsfonts,amssymb,mathtools}
\usepackage{caption}
\usepackage{subcaption}
\usepackage{multirow}
\usepackage[dvipsnames]{xcolor}

\def\tsc#1{\csdef{#1}{\textsc{\lowercase{#1}}\xspace}}
\tsc{WGM}
\tsc{QE}
\tsc{EP}
\tsc{PMS}
\tsc{BEC}
\tsc{DE}

\begin{document}
\let\WriteBookmarks\relax
\def\floatpagepagefraction{1}
\def\textpagefraction{.001}

\shorttitle{HDR ChipQA}

\shortauthors{JP Ebenezer et~al.}

\title [mode = title]{HDR-ChipQA: No-Reference Quality Assessment on High Dynamic Range Videos}

\author[1]{Joshua P. Ebenezer}

\cormark[1]

\fnmark[1]

\ead{joshuaebenezer@utexas.edu}


\address[1]{{Laboratory for Image and Video Engineering},{The University of Texas at Austin}, {Austin},{78712}, {USA}}

\author[1]{Zaixi Shang}
\fnmark[1]

\author[2]{Yongjun Wu}
\author[2]{Hai Wei}
\author[2]{Sriram Sethuraman}
\author[1]{Alan C. Bovik}

\address[2]{{Amazon Prime Video},{Seattle},{98109}, {WA},{USA}}

\cortext[cor1]{Corresponding author}

\fntext[fn1]{J.P. Ebenezer and Z. Shang contributed equally to this work}

\begin{abstract}
 We present a no-reference video quality model and algorithm that delivers standout performance for High Dynamic Range (HDR) videos, which we call HDR-ChipQA.  HDR videos represent wider ranges of luminances, details, and colors than Standard Dynamic Range (SDR) videos. The growing adoption of HDR in massively scaled video networks has driven the need for video quality assessment (VQA) algorithms that better account for distortions on HDR content. In particular, standard VQA models may fail to capture conspicuous distortions at  the extreme ends of the dynamic range, because the features that drive them may be dominated by distortions {that pervade the mid-ranges of the signal}. We introduce a new approach whereby a local expansive nonlinearity emphasizes distortions occurring at the higher and lower ends of the  {local} luma range, allowing for the definition of additional quality-aware features that are computed along a separate path. {These features are not HDR-specific, and also improve VQA on SDR video contents, albeit to a reduced degree.} {We} show that this preprocessing step significantly boosts the power of distortion-sensitive natural video statistics (NVS) features when used to predict the quality of HDR content. In similar manner, we separately compute novel wide-gamut color features using the same nonlinear processing steps. We have found that our model significantly outperforms SDR VQA algorithms on the only publicly available, comprehensive HDR database, while also attaining state-of-the-art performance on SDR content.
\end{abstract}




\maketitle


\section{Introduction}
%
%
%
%
Dynamic range is the ratio of the largest and smallest usable values of a signal. In the context of video, it is the luminance range over which a video system can capture, store, and display scenes without loss of details. Video systems and standards such as ITU BT 709~\cite{bt709} were originally designed for now-obsolete Cathode Ray Tube (CRT) displays. The BT 709 transfer function (also called the gamma curve), color space, and luminance range are still used to define how Standard Dynamic Range (SDR) video is captured, stored, and displayed.  SDR video has a luminance range of 0.1 cd/m$^2$ to 100  cd/m$^2$, while the human eye can perceive luminance levels between $10^{-6}$ cd/m$^2$ and $10^{8}$ cd/m$^2$ by deploying various mechanical, photochemical, and neuronal adaptive processes~\cite{kunkel2016perceptual}. Historically, the limitations of sensor, storage, and device technolo{g}ies constrained the luminance ranges of consumer video, but despite significant advances in sensor and display technologies, most video content is still created under SDR standards and specifications.

High Dynamic Range (HDR) videos are defined by a set of standards and technologies that enable the display of videos having larger luminance ranges than SDR. The upper limits of the allowable luminance ranges of HDR are higher than those of SDR, enabling the display of brighter highlights, while the lower limits fall below {those of} SDR, enabling the display of darker, deeper shadows. The ranges of colors that can be represented by HDR video are also broader and richer than by SDR.
\par
 HDR10 is an open HDR standard defined by the Consumer Technology Association~\cite{cta}, and is currently the most common HDR format. HDR10 videos must have depth of 10 bits and must follow the the Rec. BT 2020~\cite{bt2020}, followed by 4:2:0 color subsampling. The BT 2020 color gamut covers 75.8\% of the CIE 1931 color space, while the SDR BT 709 color gamut only covers 35.6\% of the CIE 1931 color space. The HDR10 standard allows for a maximum luminance of 10,000 nits, which is much greater than on SDR displays.
HDR10 videos must also contain static metadata describing the color volume of the mastering display, the maximum content light level, and the maximum frame-average light level. This metadata allows displays that have a lower dynamic range than the content to intelligently tonemap and adjust videos. HDR10+~\cite{hdr10plus} and Dolby Vision~\cite{dolby} are more advanced HDR standards that also include dynamic metadata, so that displays can make tonemapping and video adjustment decisions on a frame-by-frame basis.
\par 
HDR10 has seen widespread adoption in the past few years. Streaming services such as Amazon Prime Video and Netflix now support HDR10, and it is the default standard for UHD BluRays. YouTube also supports user-generated HDR10 content. Monitor and television display manufacturers like Samsung, Dell, Panasonic, and LG market HDR displays, and manufacturers such as Lenovo and Apple have released laptops with HDR displays. The increasing industry adoption of HDR and the contemporaneous creation of large amounts of high quality HDR content has accelerated adoption, and helped further enhance the Quality of Experience (QoE) of millions of viewers. However, the increased data volumes implied by HDR, along with increases of other video dimensions such as resolution and frame rate, and the growing popularity of video, stresses global networks and requires increased compression and other processing protocols. Video quality assessment (VQA) models reliably predicting HDR video quality as they are processed are necessary for monitoring, mediating, and controlling high-volume HDR video traffic. Effective perceptual VQA tools can significantly impact both bandwidth consumption and customer satisfaction. Today, VQA models are used to automatically perceptually optimize per-frame, per-shot, and per-title bitrate ladder decisions at very large scales of commercial deployment. However, advancements on true HDR video quality modeling and algorithm design remains quite limited.
\par 
The gold standard of video quality measurements are the subjective opinions of a suitable population of viewers. Of course, collecting such subjective scores for videos is an expensive and time-consuming task and cannot be done at the scales of video streaming services. However, if a large enough set of representative HDR videos were collected, with representative distortions, and scored by a large enough group of viewers, then the collected scores could be used to train objective VQA algorithms suitable for deployments at scale.
\par 
VQA algorithms are categorized as Full-Reference (FR) or No-Reference (NR). NR VQA algorithms do not use any information from the pristine reference versions of the distorted videos, unlike FR VQA algorithms, and instead attempt to predict the quality of distorted videos without additional input. NR VQA is a much harder problem than FR VQA, but is one that needs a solution since quite commonly no reference video is available. While effective FR and NR VQA models exist for SDR content, naive application of these to HDR content may not yield the best results, as we will show. Here we show how perceptually-driven changes can be made to an existing high-performance SDR NR VQA algorithm to improve its quality prediction power on HDR video content.

\section{Related Work}

 Most existing NR VQA models implicitly assume that their inputs are luma, and not luminance, since images and videos are digitally stored as luma {and color-difference channels} and not luminance. We will not process luminance values in our proposed models, and wherever we refer to HDR luma, {color-difference channels}, or R$^\prime$G$^\prime$B$^\prime$, we mean RGB light intensity values that have been passed through the PQ OETF and then converted to luma and {color-difference channels} through the linear transformation defined in BT 2020. 
\par 
To the best of our knowledge, there do not exist any NR VQA models that have been shown to yield enhanced performance on true HDR. Current NR Image Quality Assessment (IQA) algorithms, which do not measure temporal aspects of videos, can be applied on a frame-by-frame basis. Frame-based methods are effective enough in the absence of temporal distortions, high motions, or quality variations over time. It has been shown~\cite{videval,pooling} that this is often the case for user-generated content, on which NR IQA frame algorithms have been found to deliver state-of-the-art predictive power. BRISQUE~\cite{brisque} is an NR IQA algorithm that is based on natural scene statistics (NSS) models, exploiting the fact that images and videos reliably obey statistical regularities in the absence of distortions~\cite{ruderman}, and that the visual brain has adapted to these statistics to process visual stimuli in an efficient way. 
\par 
For example, the statistics of Mean-Subtracted Contrast-Normalized (MSCN) coefficients of luma are expressed using NSS models to quantify image quality in BRISQUE, and in a related unsupervised model called NIQE~\cite{niqe}. HIGRADE~\cite{higrade}, which was originally designed for tonemapped (8-bit) picture IQA, models the statistics of the gradient and log-derivative of each channel in the CIELAB~\cite{cielab} color space.
\par
V-BLIINDS~\cite{vbliinds} is an NR VQA algorithm that models the statistics of the discrete cosine transforms of frame differences, and is an early example of using natural video statistics (NVS) models as the basis for conducting video quality prediction. It also contains features that capture various aspects of motion. ChipQA~\cite{chipqa} is a more recent VQA algorithm that models the statistics of space-time chips, which are highly localized space-time slices of MSCN frames. An optimal space-time direction along which to extract quality-aware features is chosen by finding an oriented chip having minimum excess kurtosis. ChipQA also contains color and gradient features.
\par
RAPIQUE~\cite{rapique} is an NR VQA algorithm designed to conduct quality analysis on SDR User Generated Content (UGC). RAPIQUE contains various NSS features and features extracted using a Convolutional Neural Network (CNN). All of the features are pooled over time and used to train a support vector regressor.
 HDR-BVQM~\cite{hdrbvqm} consists of BRISQUE features, the log-derivative features defined from HIGRADE, and the temporal features from VBLIINDS. All of the HDR-BVQM features were designed for SDR videos, with no adjustment being made to specify them for application to HDR. Hence, it is really an SDR NR VQA algorithm, although it has been tested on HDR content.
VSFA~\cite{vsfa} is an NR VQA algorithm that consists of a pre-trained CNN, a Gated Recurrent Unit (GRU) network, and a temporal pooling layer. {VSFA achieves state-of-the-art  NR quality prediction performance on UGC datasets. NorVDPNet~\cite{norvdpnet} is an NR IQA method intended for HDR. It utilizes a 2D CNN network that is trained on HDR VDP~\cite{hdrvdp} scores between reference and distorted image pairs. NorVDPNet treats HDR VDP scores as a proxy for quality scores.}
 \par 
TLVQM~\cite{tlvqm} is an NR algorithm that uses a large number of hand-designed, low complexity features that are specific to commonly occurring distortions like blockiness, blurring, motion artifacts, jerkiness, interlacing, and so on. Other `high-complexity features' are sampled at 1 Hz to capture sharpness, blockiness, noise, color, and contrast. The HCF set are defined on the CIELAB space, which was designed for SDR. All of the models ChipQA, HIGRADE, and TLVQM are defined using the CIELAB color space, which was designed for SDR content. Later, when we conduct experimental studies of these and other models on HDR content, we will instead implement them using the HDR-CIELAB~\cite{hdrcielab} space instead of the CIELAB space to improve their performances for fairer comparison.

\section{Proposed Video Quality Assessment algorithm}

Our new HDR VQA algorithm has 3 parts: a spatial luma component, a spatio-temporal luma component, and a color component. We designed the model to extend the SDR BRISQUE and ChipQA algorithms by redefining their features sets to produce enhanced performance on the HDR VQA problem. A key ingredient of our design is the introduction of a non-linear processing stage that sensitizes the model to distortions affecting the high and low ends of the dynamic range.

\subsection{Spatial Luma Component}
~\label{subsection:Luma}
We begin by explaining the way quality-aware spatial features are defined and computed {on} each frame of a video to be analyzed.
\subsubsection{Statistical features.}

The MSCN coefficients $\hat{V}\lbrack i,j,k_0 \rbrack$ of a luma or {R$^\prime$G$^\prime$B$^\prime$  channel} of a video frame $V\lbrack i,j,k_0 \rbrack$ are defined as :
\begin{equation}\label{eq:mscn} 
\hat{V}\lbrack i,j ,k_0 \rbrack = \frac{V\lbrack i,j,k_0 \rbrack-\mu\lbrack i,j,k_0 \rbrack}{\sigma\lbrack i,j ,k_0 \rbrack +C},    
\end{equation}
where $i\in 1,2..,M$, $j=1,2..,N$ are spatial indices, {$k_0$ is the frame index}, $M$ and $N$ are the frame height and width, respectively, the constant $C$ imparts numerical stability, and where
  \begin{equation}\label{eq:mean} 
  \mu\lbrack i,j,k_0 \rbrack = \sum\limits_{k=-L}^{k=L} \sum\limits_{k=-L}^{k=L} w\lbrack m,l,k_0 \rbrack V\lbrack i+m,j+l,k_0 \rbrack
  \end{equation}
  and
\begin{multline*}\label{eq:sigma} 
\sigma\lbrack i,j,k_0 \rbrack =\Bigl( \sum\limits_{k=-L}^{k=L} \sum\limits_{k=-L}^{k=L} w\lbrack m,l,k_0 \rbrack  (V\lbrack i+m,j+l, \\ k_0 \rbrack - \mu\lbrack i,j,k_0 \rbrack )^2 \Bigr)^\frac{1}{2}
\end{multline*}
are the local weighted spatial mean and standard deviation of luma (rms contrast), respectively. The weights $w=\{w[m,l]|m=-L,\dots,L,l=-L,\dots,L\}$ are a 2D circularly-symmetric Gaussian weighting function sampled out to 3 standard deviations and rescaled to unit volume, and $K=L=3$. The MSCN coefficients (\ref{eq:mscn}) of pristine SDR videos have been observed to reliably follow a Gaussian distribution~\cite{ruderman,brisque,niqe}. The MSCN coefficients of real-world distorted SDR videos predictably depart from Gaussianity, but can often be effectively modelled as following Generalized Gaussian Distributions (GGD), defined as

  \begin{equation}
  g_1(x;\alpha;\beta) = \frac{\alpha}{2\beta \Gamma(\frac{1}{\alpha})} \exp [-(\frac{|x|}{\beta})^\alpha]
  \end{equation}\label{ggd}
where $\Gamma(.)$ is the gamma function:
\begin{equation}
\Gamma(\alpha) = \int_0^\infty t^{\alpha-1} \exp(-t) dt.
\end{equation}
The estimated shape parameter $\alpha$ and variance $\beta$ are estimated and used as features for quality assessment. Statistical regularities of distorted and naturalistic pictures and videos have been observed on SDR content, and it is reasonable to suggest that they should be applicable to HDR videos as well, albeit, with suitable modifications and/or additional processing steps. HDR10 has been designed to produce more accurate and perceptually correct representations of the natural world than SDR content, and thus might be expected to reliably obey appropriate NSS and NVS models.
 
As expected, we have found that the MSCN coefficients of pristine HDR luma frames can be reliably modeled as approximately following a Gaussian probability law. To show this, following~\cite{strred}, we computed the ratio of the difference between the entropy of the fitted Gaussian distribution and the entropy of the empirical distribution (denoted by $\Delta H$), to the entropy of the empirical distribution (denoted by $H$) on each frame {of each} reference video in our dataset.  We found that the average value of $\frac{\Delta H}{H}$ was 0.0467, indicating that $\Delta H$ is a small fraction of $H$ for nearly all the pristine reference videos. Likewise, the statistics of HDR content subjected to distortion tend to depart from Gaussianity. The average value of $\frac{\Delta H}{H}$ on all frames of the videos that were downsized to 1080p and compressed at 1 Mbps, for example, was 0.2566.

\par
\begin{figure*}
\centering
\subfloat[]{{\includegraphics[width=0.24\textwidth]{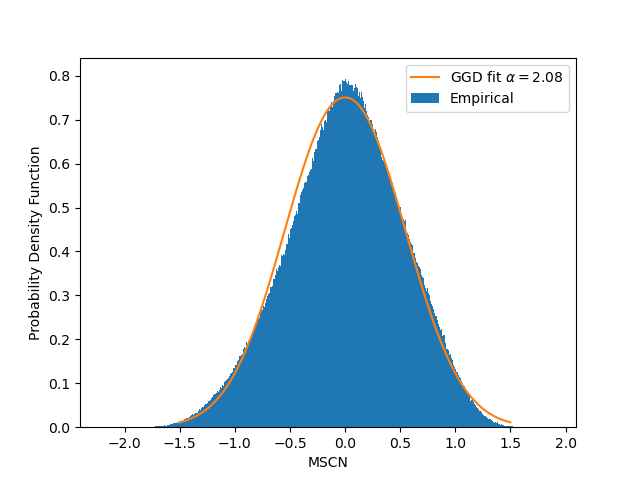}}}
\subfloat[]{{\includegraphics[width=0.24\textwidth]{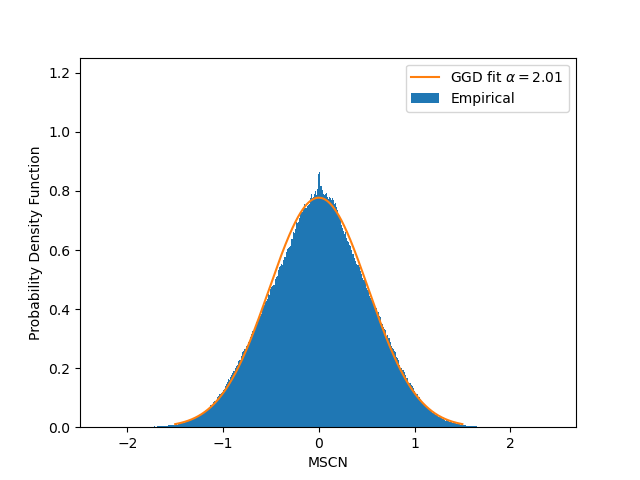}}} 
\subfloat[]{{\includegraphics[width=0.24\textwidth]{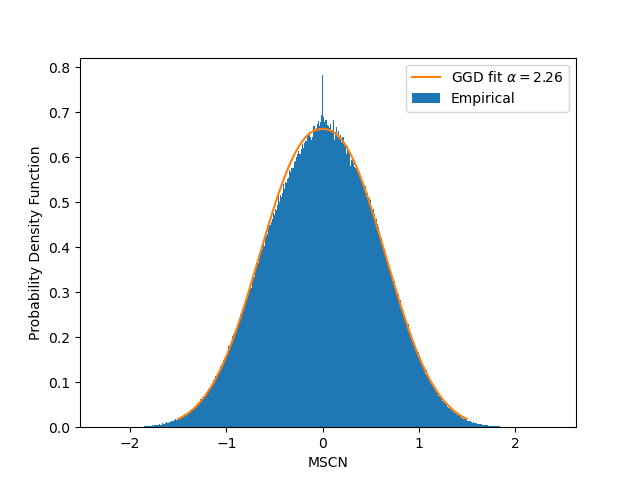}}}
\subfloat[]{{\includegraphics[width=0.24\textwidth]{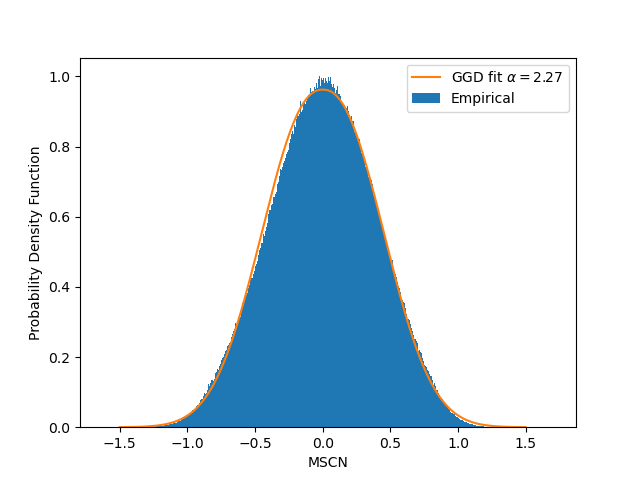}}}
\caption{Empirical distributions of MSCN coefficients of pristine HDR frames (blue) with best GGD fits superimposed (orange).  }
\label{fig:pristine_luma_mscn}
\end{figure*}

 
By way of example, consider the histograms of the MSCN coefficients of randomly selected pristine PQ coded HDR video frames, computed from (\ref{eq:mscn}) with $C=4$, as shown in Fig.~\ref{fig:pristine_luma_mscn}, along with their GGD fits, where the values of the best-fit shape parameters are given in the legend. As may be observed, the shape parameters of the GGD fits are close to 2, indicating that they near Gaussianity. The value of $C$ {that is used to compute the} MSCN coefficients of SDR content (which have a digital range of 0-255) is typically taken to be 1~\cite{brisque,niqe}. However, HDR content has a digital range of 0-1023, a four-fold increase of the digital range, hence we used $C=4$ {when} computing {the} MSCN coefficients of HDR content. \par
To capture the correlation structure of spatial frames, we compute the products of neighboring pairs of pixels, as in BRISQUE, and these are modelled as following an Asymmetric Generalized Gaussian Distribution (AGGD). The parameters of the best AGGD fit to the histograms of each of the pairwise products are extracted and used as quality-aware features.

 
\subsubsection{Nonlinear expansion of luma/color extremes}
\label{brisque_fx}

{In order to sensitize our model to distortions that are particular to HDR and wide color gamut (WCG) content, we developed a method that applies an expansive nonlinearity to enhance the ends of the brightness and color scales. It turns out that this approach is remarkably effective.} 

\begin{figure}
\centering
\begin{subfigure}{0.23\textwidth}
\includegraphics[width=\textwidth]{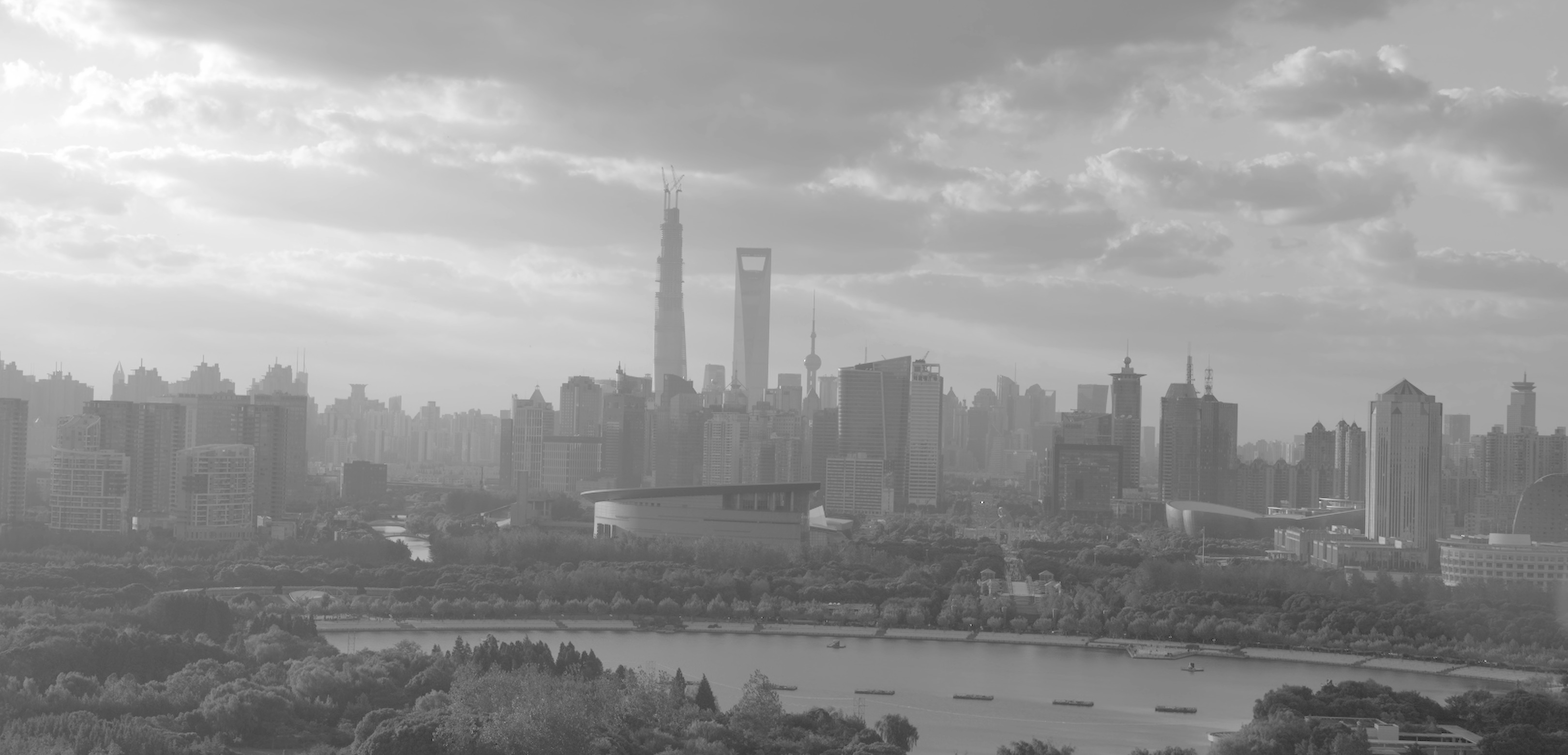}
\caption{HDR version}
\label{fig:hdrpristine}
\end{subfigure}
\begin{subfigure}{0.23\textwidth}
\includegraphics[width=\textwidth]{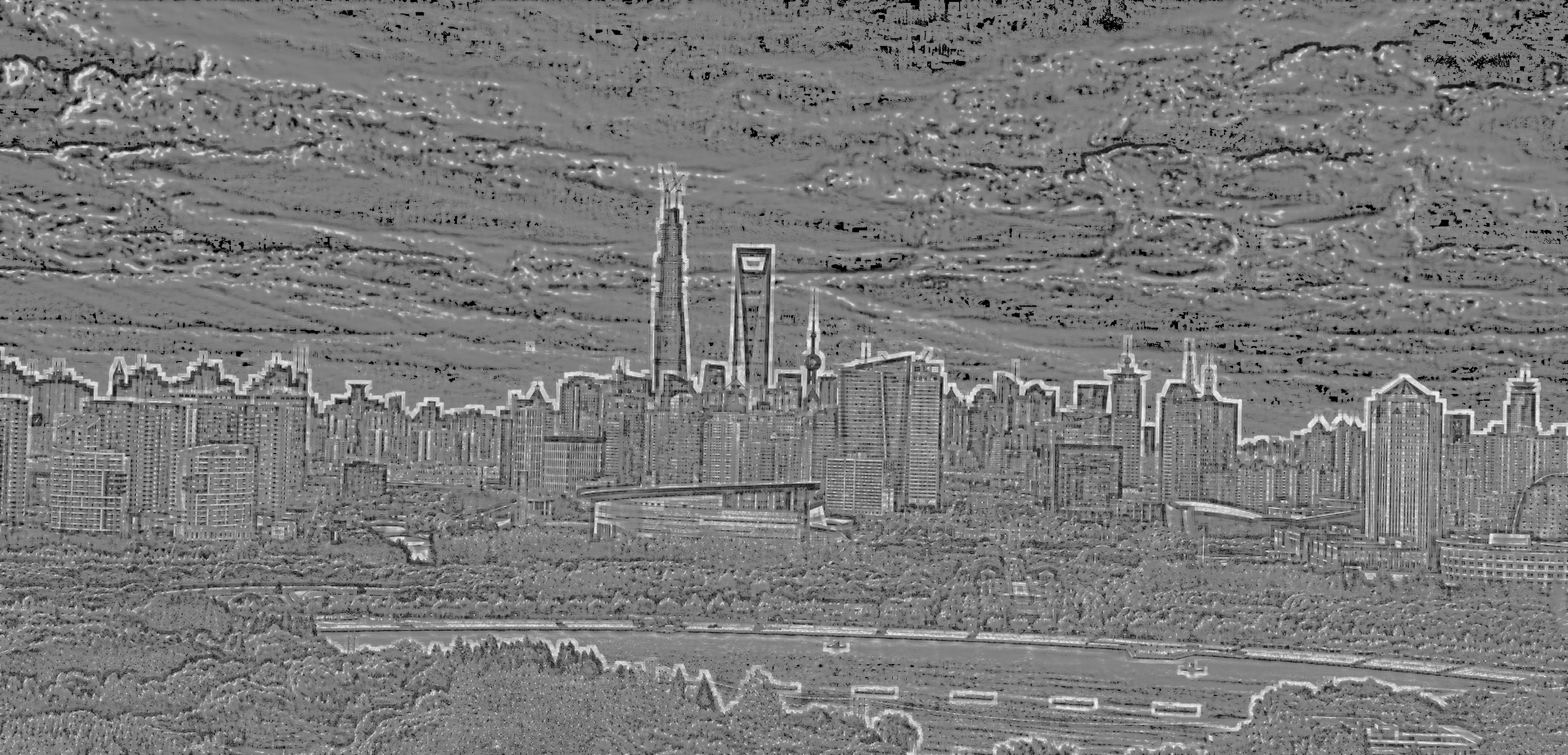}
\caption{HDR version after  $f(x)$}
\label{fig:hdrpristinelnl}
\end{subfigure}\\
\begin{subfigure}{0.23\textwidth}
\includegraphics[width=\textwidth]{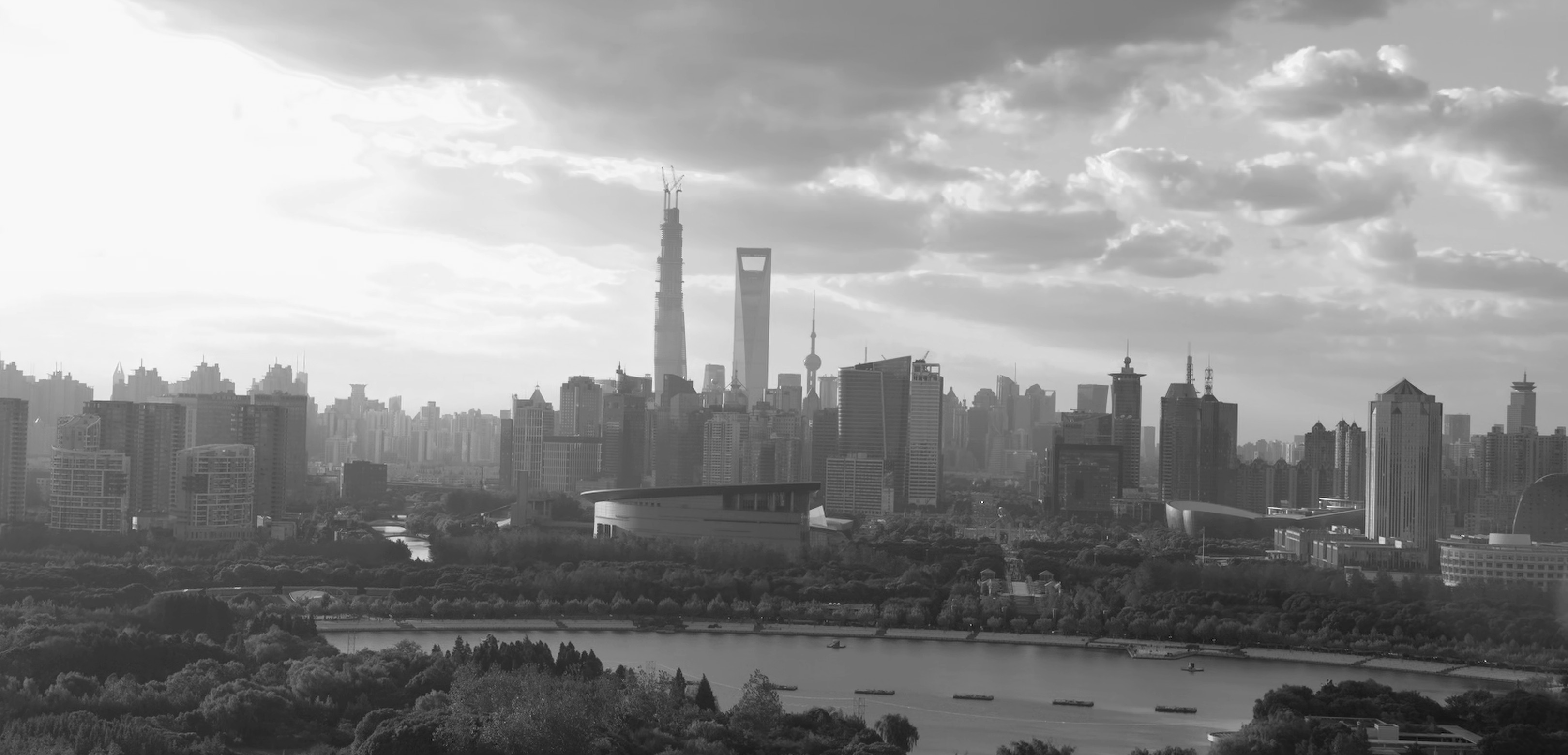}
\caption{SDR version}
\label{fig:sdrpristine}
\end{subfigure}
\begin{subfigure}{0.23\textwidth}
\includegraphics[width=\textwidth]{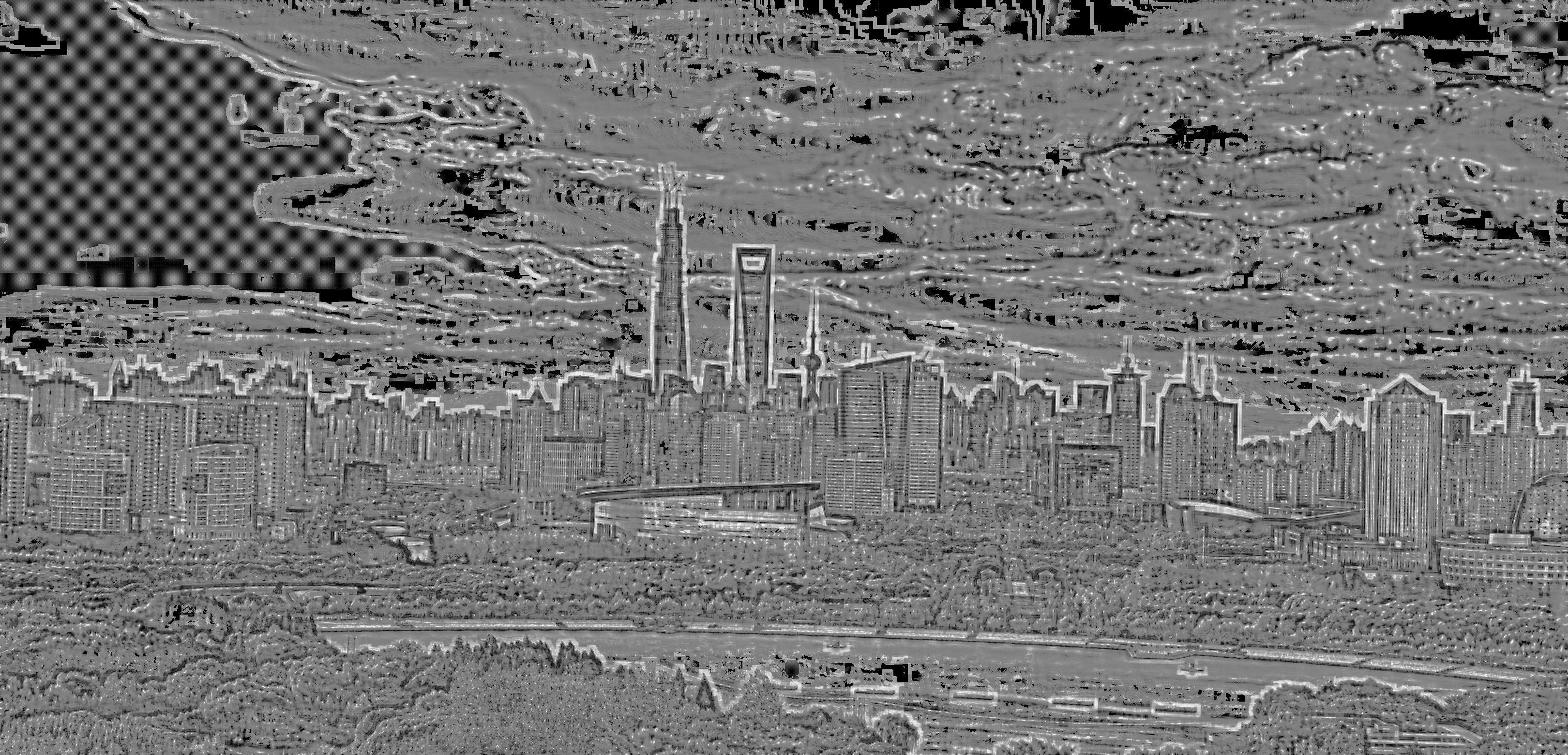}
\caption{SDR version after  $f(x)$}
\label{fig:sdrpristinelnl}
\end{subfigure}
\caption{Original frames (left) and nonlinearly transformed frames (right) of ``Cloud" video using (\ref{eq:exp_piecewise}) with $\delta=4$.}
\label{fig:clouds}
\end{figure}

{Due to its higher bit depth, HDR excels in representing local contrasts and details without causing saturation of luma or color. For example, as shown in Fig.~\ref{fig:clouds}, a video of a bright sky may appear saturated and overexposed and will therefore lack details in SDR, while the greater dynamic range of HDR may allow the delineation of clouds and other details in the same sky. Likewise, distortions occurring on very bright and/or very dark regions (of HDR content) which might not be perceivable in SDR, can become obvious on HDR.}  

We have observed that when true HDR/WCG video signals are distorted by common artifacts arising from compression and transmission, they are {still} effectively handled by existing VQA models (those designed on SDR). However, the feature responses to HDR-enabled details in bright and dark regions could be masked by the feature responses to regions of the frame that can be represented in both SDR and HDR. Because of this, while distortions on HDR-enabled regions of video frames can be singularly annoying, {existing} VQA features computed on them may be highly diluted by those computed on other spatial areas of HDR video frames. This may be the principal reason why standard VQA models originally designed for SDR contents do not perform very well on HDR video signals.


Our innovation in this regard is to introduce a separate parallel, feature computation pathway, which deploys a simple expansive point nonlinearity prior to the computation of quality-aware statistical features. This is different from the usual forward modeling of the flow of visual information along the visual pathway {that} we, and others, often have used to define and compute quality-sensitive video features, {but prior approaches have failed to adequately access the perceptual effects of HDR distortions, since they are dominated by distortions on regions not expressing brightnesses and colors at the extremes, but usually covering much larger spatial image geographies. By introducing an expansive nonlinearity,} the extreme ends of the dynamic ranges of brightness and color are spread, while the middle ranges that dominate standard (SDR-like) feature computations {are compressed}. Statistical aspects of distortions on regions of extreme local luma/color that would have not been accounted for because they generally occupy much smaller spatial regions, become dominant by preferentially enhancing them in a separate, non-linearly transformed feature channel.
\par
The expansive nonlinearity is applied as follows. {First, at every spatial frame coordinate, define a window or patch of size $W\times W$, within which the sample values (luma or color) are linearly mapped to} $[-1,1]$. {While the exact range of the linear mapping applied prior to the nonlinearity is not significant, using $[-1,1]$ produces symmetric results that are conceptually and algorithmically simpler}. {Note that the $W\times W$ windows are heavily overlapping.} Later, we will explain the effects of varying the patch dimension $W$ and the relative merits of mapping entire frames to $[-1,1]$, rather than patches.

\begin{figure}[h]
\centering
  \includegraphics[width=0.3\textwidth]{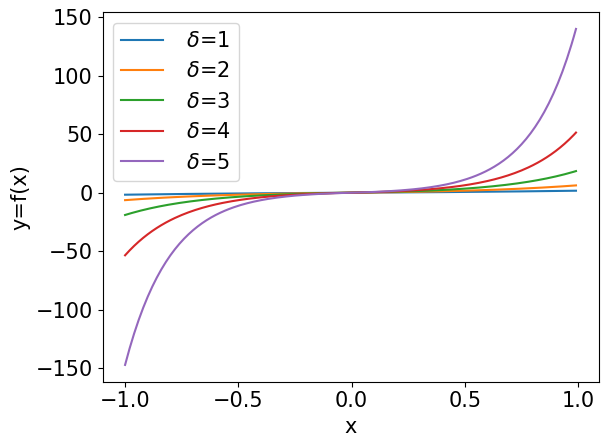}%
  \caption{Plot of expansive nonlinearity in (\ref{eq:exp_piecewise}) for five values of the expansion parameter $\delta$}
  \label{fig:exp_piecewise}
\end{figure}

Then pass the luma or {R$^\prime$G$^\prime$B$^\prime$  values of each locally-mapped} frame through the expansive nonlinearity 
\begin{equation}
\label{eq:exp_piecewise}
        f(x;\delta) =   \begin{cases} \exp(\delta x)-1 & x>0 \\
    1-\exp((-\delta x)) & x<0 \end{cases}
\end{equation}
which is piecewise monotonic, and plotted in Fig.~\ref{fig:exp_piecewise} for $\delta=1,2,3,4,$ and $5$. {This simple nonlinearity has the effect of expanding the extreme ends of the ranges of luma or color, while compressing these ranges away from the extremes. As we will show,} extracting statistical features on videos that have been ``HDR sensitized" by applying (\ref{eq:exp_piecewise}) produces much higher correlations against subjective quality judgments. Of course, (\ref{eq:exp_piecewise}) is applied on the rescaled luma {and R$^\prime$G$^\prime$B$^\prime$ } values, which are themselves obtained by applying the PQ OETF on {RGB light intensity values.} 
\par

  \begin{figure}
\centering
\begin{subfigure}{0.23\textwidth}
\includegraphics[width=\textwidth]{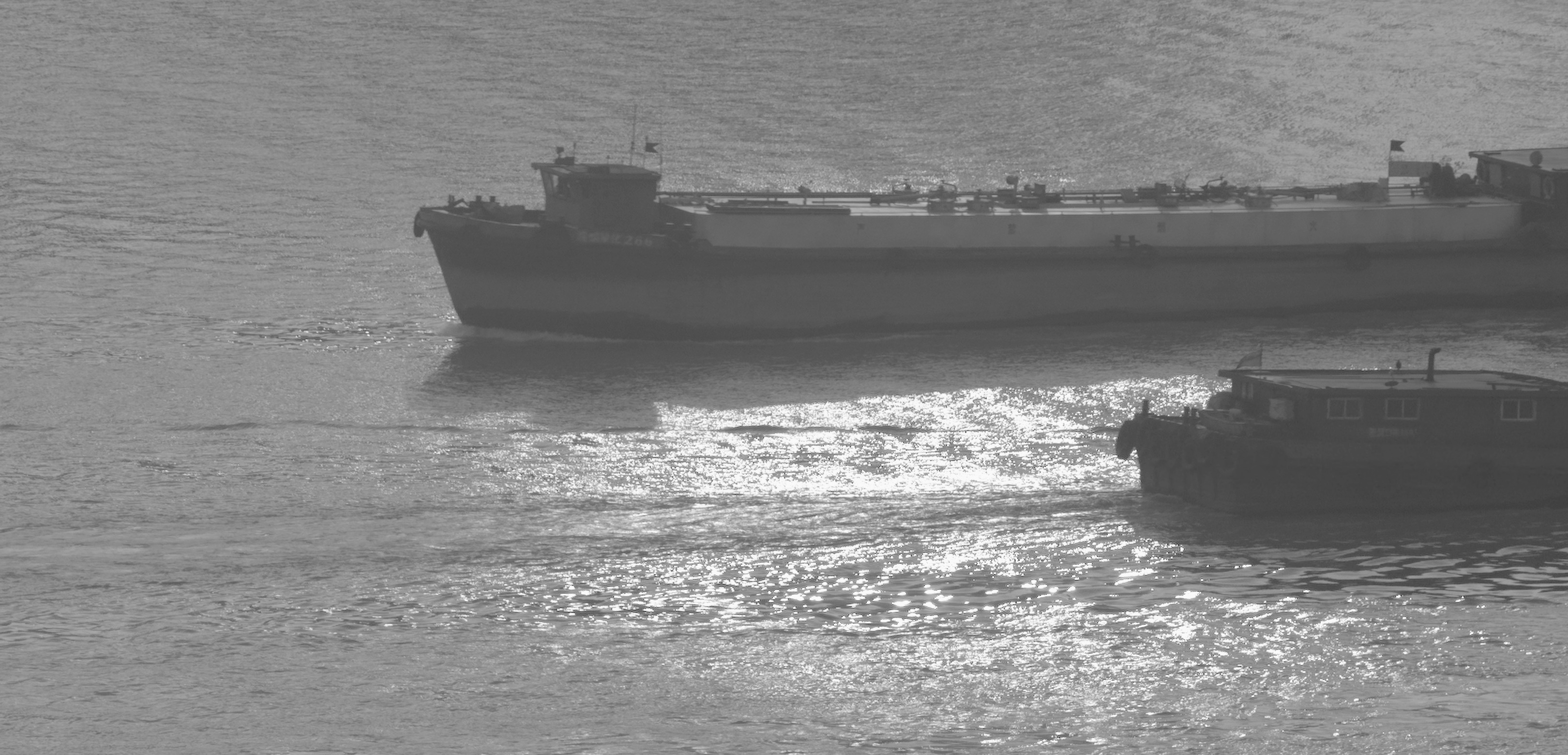}
\caption{Pristine}
\label{fig:boatpristine}
\end{subfigure}
\begin{subfigure}{0.23\textwidth}
\includegraphics[width=\textwidth]{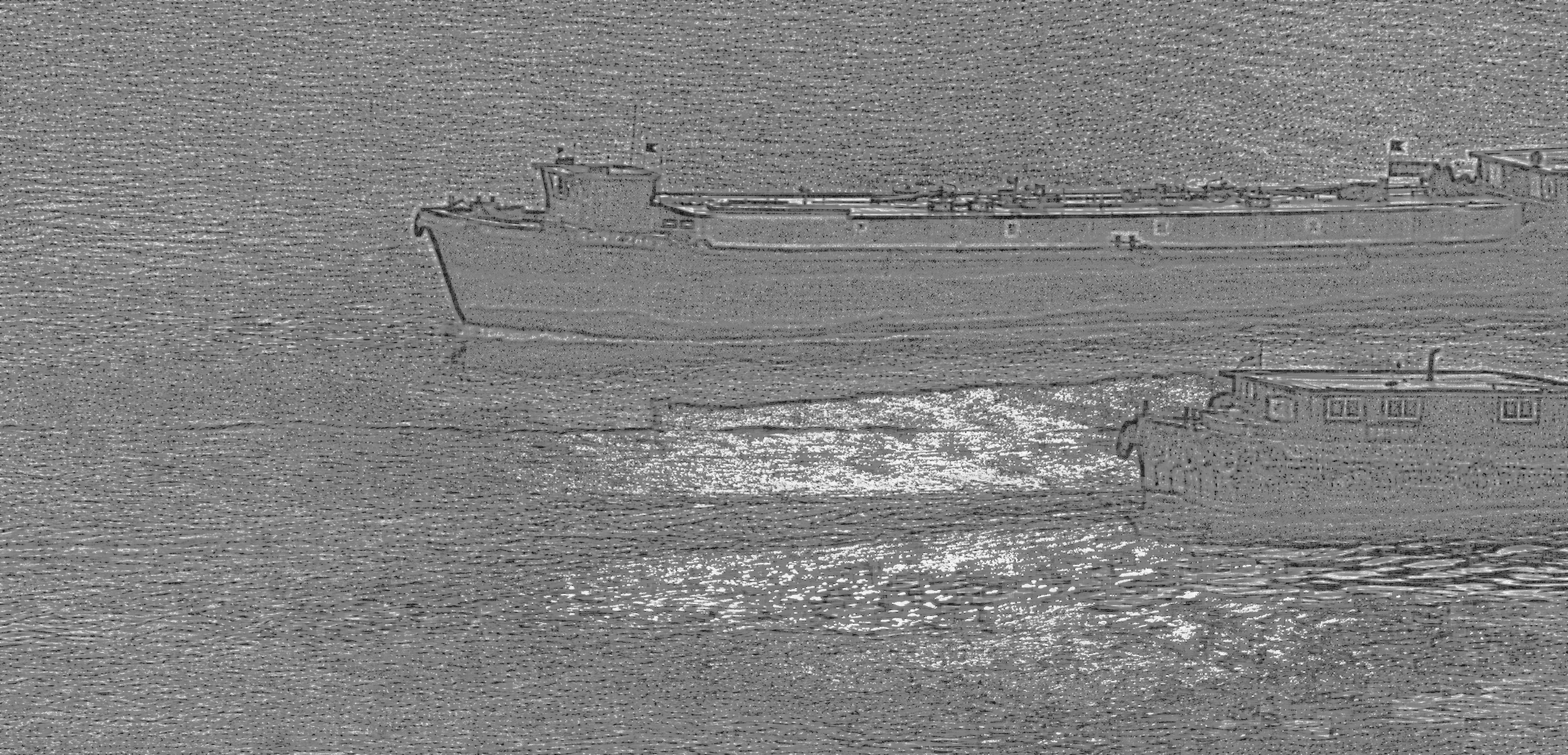}
\caption{Pristine after  $f(x)$}
\label{fig:boatpristine_lnl}
\end{subfigure}\\
\begin{subfigure}{0.23\textwidth}
\includegraphics[width=\textwidth]{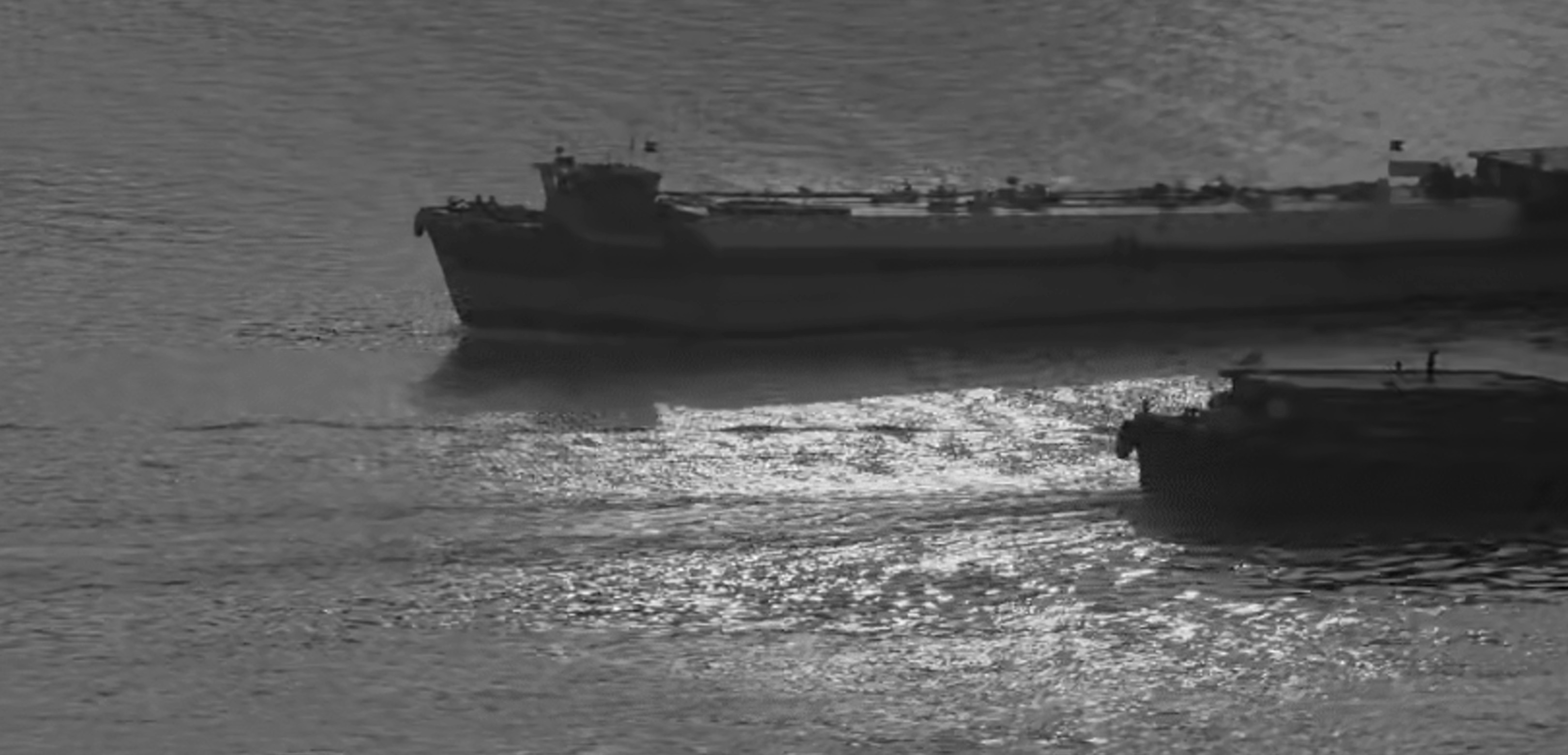}
\caption{Distorted}
\label{fig:boatdistorted}
\end{subfigure}
\begin{subfigure}{0.23\textwidth}
\includegraphics[width=\textwidth]{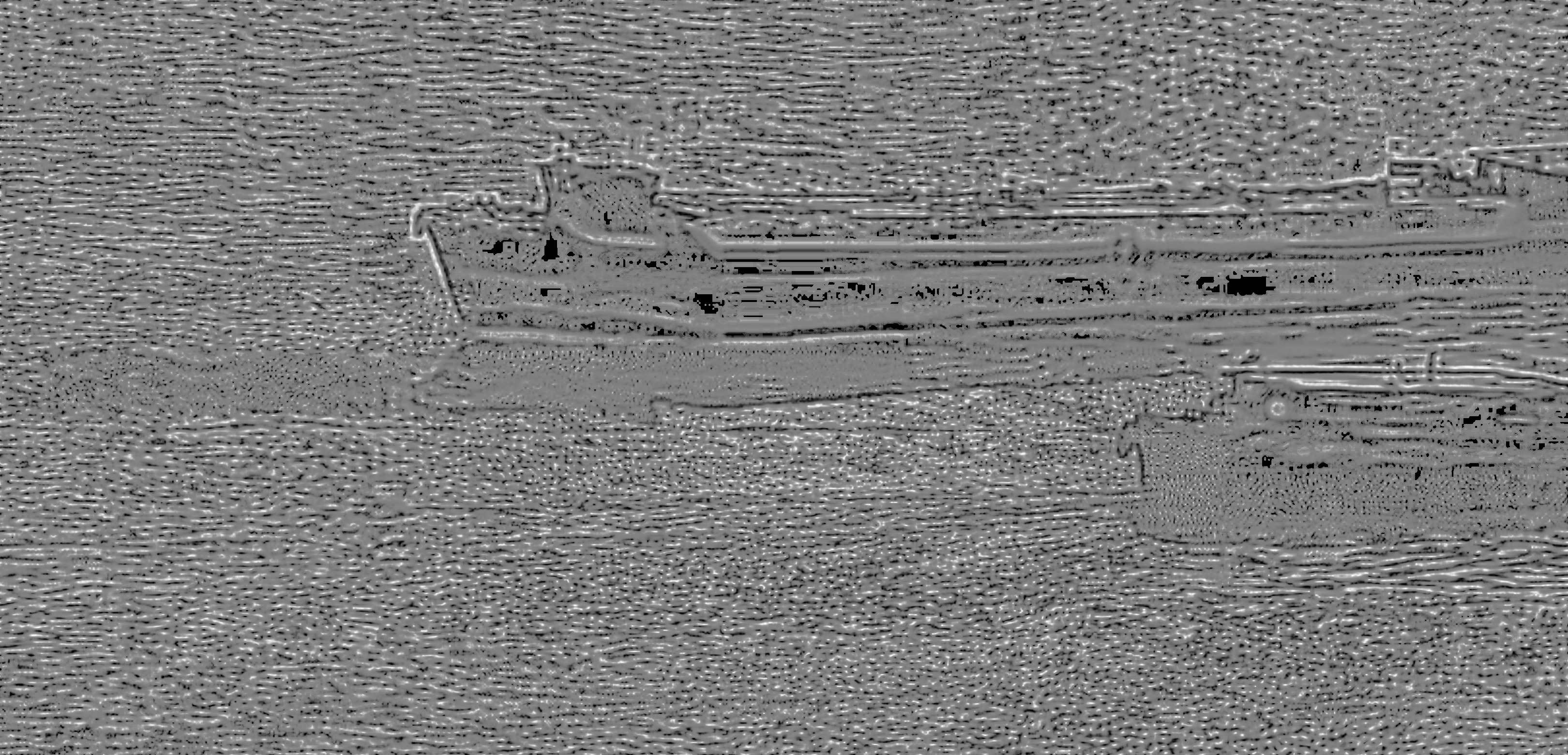}
\caption{Distorted after  $f(x)$}
\label{fig:boatdistortedlnl}
\end{subfigure}
\caption{Original frame (left) and nonlinearly transformed frame (right) of ``Ocean" video using (\ref{eq:exp_piecewise}) with $\delta=4$.}
\label{fig:display_nl_boat}
\end{figure}

  \begin{figure}
\centering
\begin{subfigure}{0.23\textwidth}
\includegraphics[width=\textwidth]{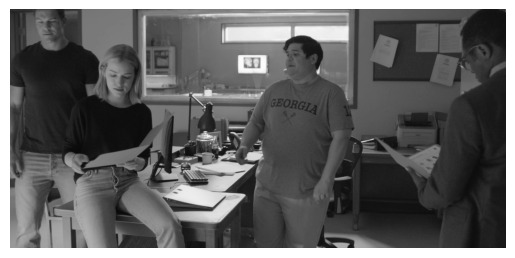}
\caption{Pristine}
\label{fig:rchrpristine}
\end{subfigure}
\begin{subfigure}{0.23\textwidth}
\includegraphics[width=\textwidth]{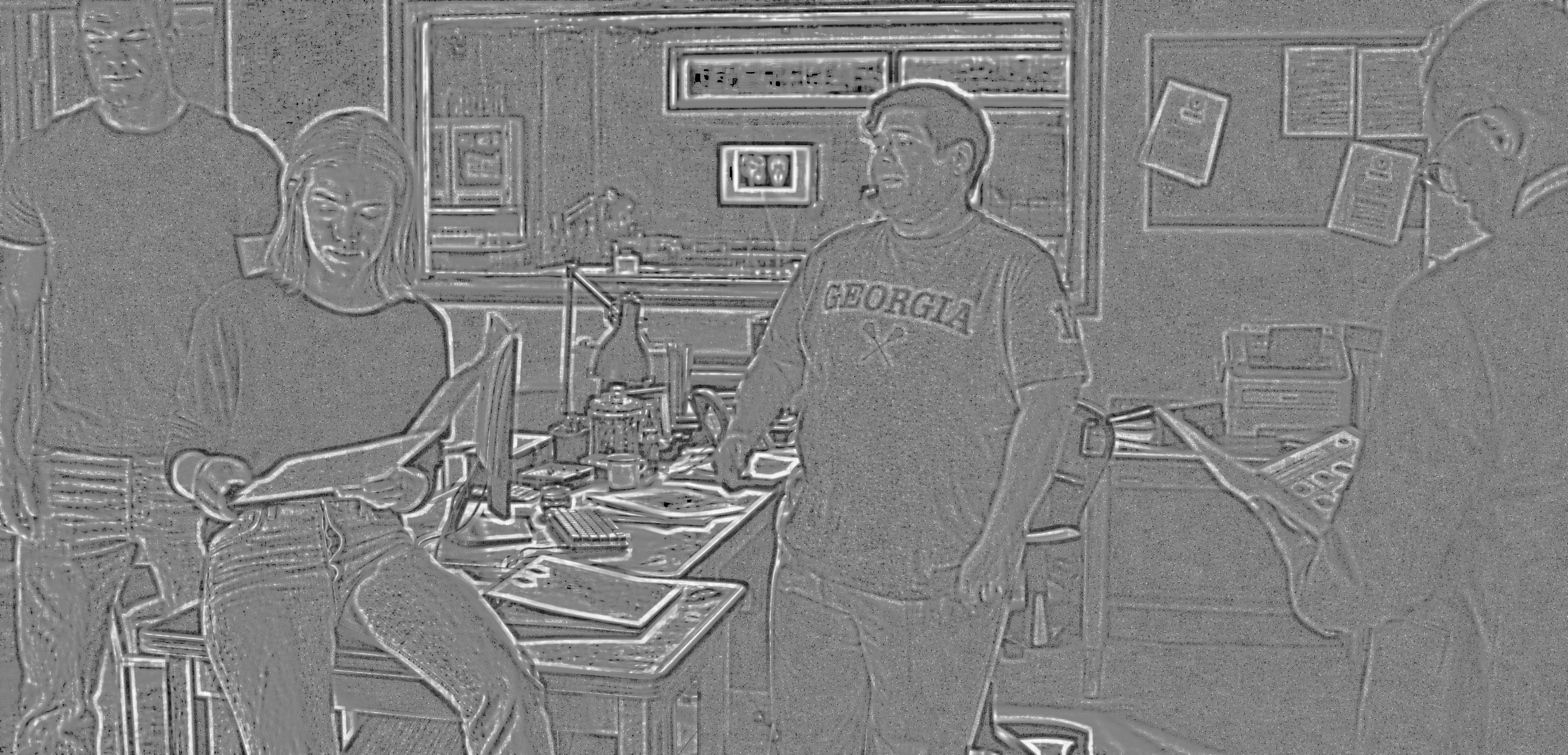}
\caption{Pristine after  $f(x)$}
\label{fig:rchrpristine_lnl}
\end{subfigure}\\
\begin{subfigure}{0.23\textwidth}
\includegraphics[width=\textwidth]{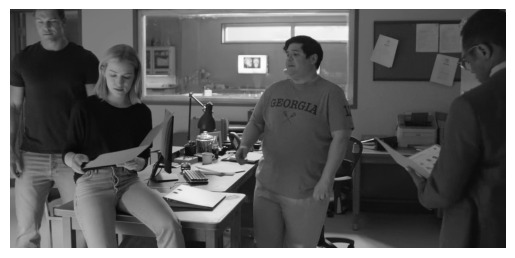}
\caption{Distorted}
\label{fig:rchrdistorted}
\end{subfigure}
\begin{subfigure}{0.23\textwidth}
\includegraphics[width=\textwidth]{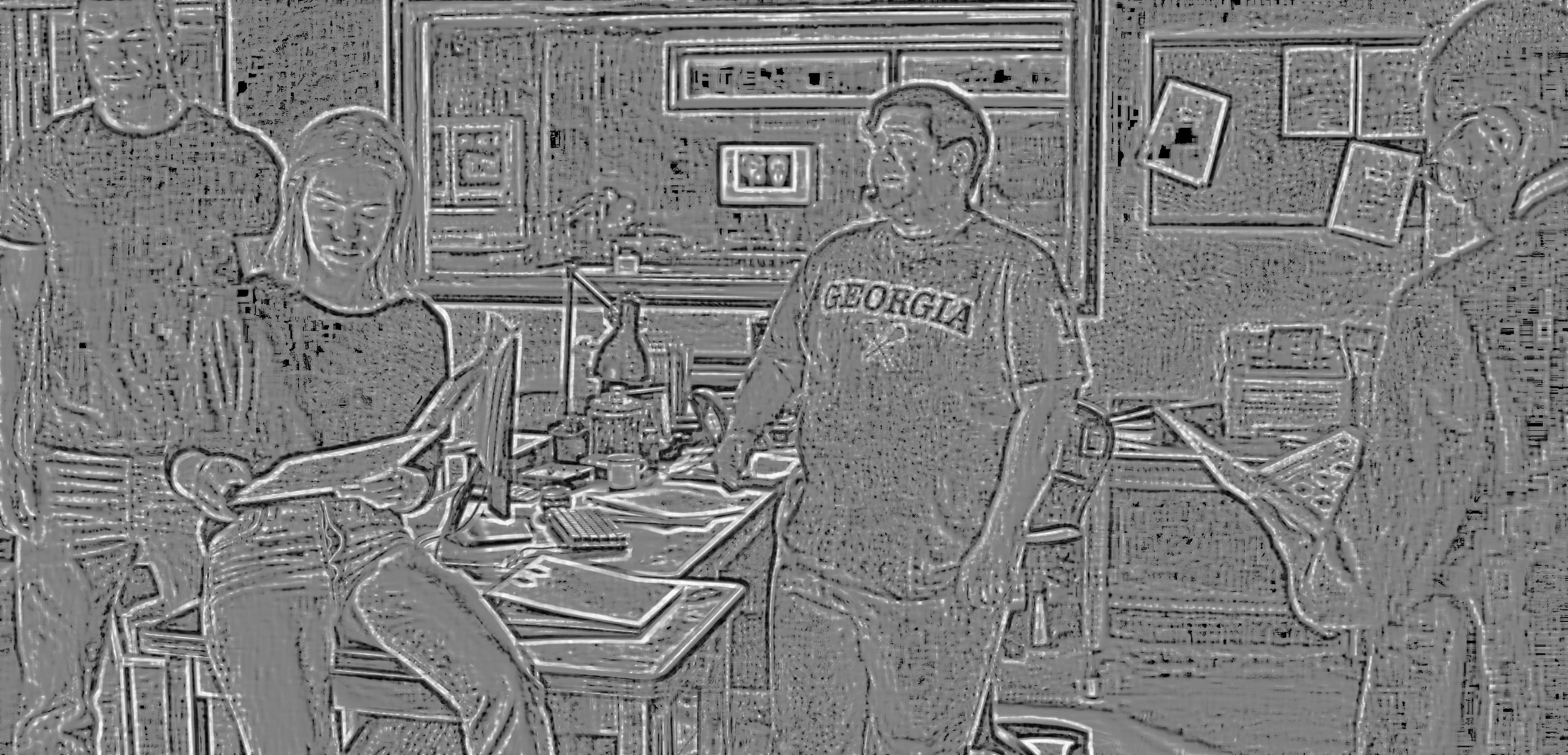}
\caption{Distorted after  $f(x)$}
\label{fig:rchrdistortedlnl}
\end{subfigure}
\caption{Original frame (left) and nonlinearly transformed frame (right) of ``Reacher" video using (\ref{eq:exp_piecewise}) with $\delta=4$.}
\label{fig:display_nl_rchr}
\end{figure}

Figs.~\ref{fig:hdrpristinelnl} and \ref{fig:sdrpristinelnl} show the result of $f(x)$ applied on HDR and SDR frames. $f(x)$ enhances the local contrast and details that the HDR version is able to represent in the cloudy region in the top left of the frame, which the SDR region is unable to represent due to its limited bit-depth. This enables feature responses to be more attuned to such regions and prevents masking from the feature responses to other regions.
\par
Figs.~\ref{fig:display_nl_boat} and ~\ref{fig:display_nl_rchr} depict examples of HDR luma frames (tonemapped for viewing in this paper) before and after applying $f(x;4)$  (using $\delta=4$ in (\ref{eq:exp_piecewise})) on windows of linear dimensions $W=31$. As may be seen, details near the extremes of the local luma  range {that may not have been as visible in SDR content due to its limited ability to represent details in bright or dark regions,} are amplified, while those in the mid-range of {local luma} are suppressed. In particular, in Fig~\ref{fig:boatpristine}, details are visible in the sun's reflection on the water in the pristine version, and $f(x)$ enhances the local contrast in that region and hence highlights it in Fig.~\ref{fig:boatpristine_lnl}. In the distorted version shown in Fig~\ref{fig:boatdistorted}, local contrast and details are suppressed due to compression and downscaling. $f(x)$ highlights this in Fig.~\ref{fig:boatdistortedlnl}, where the sun's reflection on the water does not stand out as much as it does in Fig.~\ref{fig:boatpristine_lnl}. The statistics and feature responses to these two images will accordingly differ.
\par 
Similarly, Fig.~\ref{fig:rchrdistorted} shows a compressed and downscaled version of Fig.~\ref{fig:rchrpristine}. Blocking and downscaling artifacts in dark regions of the distorted version, such as on the T-shirts and coats, become more visible after $f(x)$ is applied, as can be seen by comparing Fig.~\ref{fig:boatpristine_lnl} and Fig.~\ref{fig:boatdistortedlnl}. Feature responses to distortions in such dark regions would otherwise be masked by feature responses to other areas.

Thus,~(\ref{eq:exp_piecewise}) functions to highlight distortions that occur in ranges of  luma and color regions that are made possible by HDR. {Increasing the value of $\delta$ in (\ref{eq:exp_piecewise}) crushes larger ranges of pixel values while {locally} expanding narrower, more extreme ranges of the luma and color ranges. As the ranges of expansion are narrowed, the degrees of expansion are increased. However, very high values of $\delta$ should be avoided, since then the ranges of expansion near the endpoints of $[-1,1]$  become extremely narrow and excessively amplified, reducing information relevant to HDR. On the other hand, very low values of $\delta$ may not have a significant effect on the video and fail to highlight distortions at extreme ends of the luma range.}  
 \par
\begin{figure*}
\centering
\subfloat[]{{\includegraphics[width=0.24\textwidth]{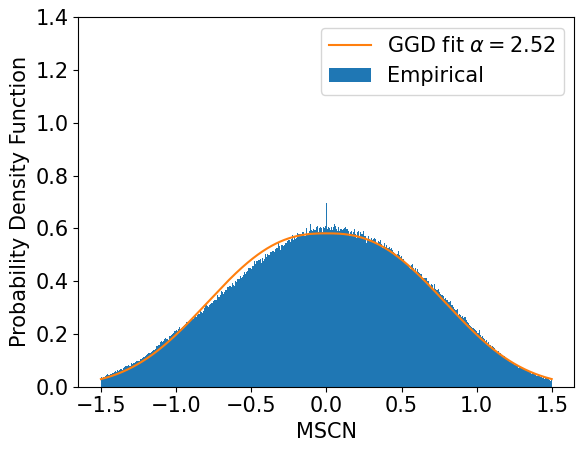}}}
\subfloat[]{{\includegraphics[width=0.24\textwidth]{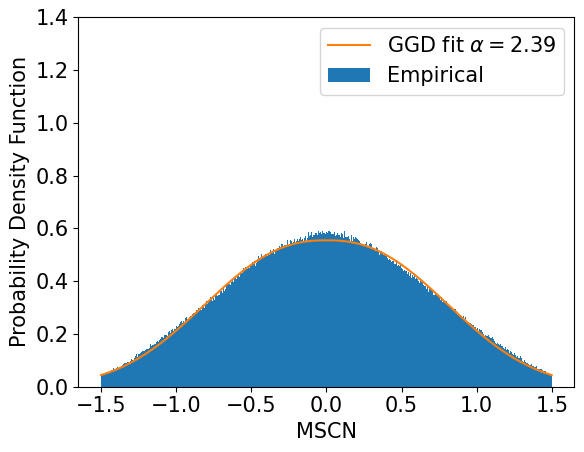}}} 
\subfloat[]{{\includegraphics[width=0.24\textwidth]{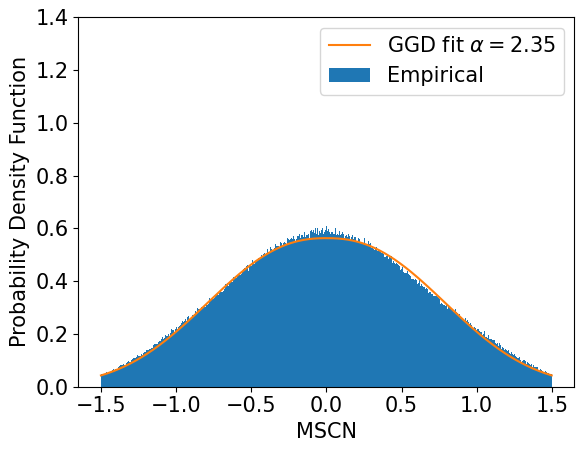}}}
\subfloat[]{{\includegraphics[width=0.24\textwidth]{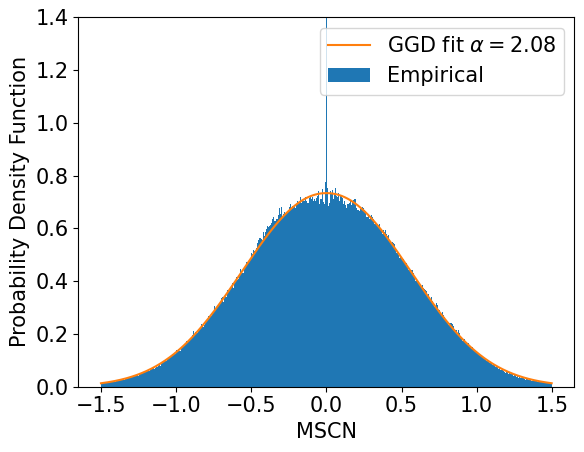}}}
\caption{Empirical distributions of MSCN coefficients of pristine HDR frames (blue) after $f(x;4)$ ((\ref{eq:exp_piecewise}) with $\delta=4$) was applied along with their best GGD fits (orange). }
\label{fig:exp_ggd_ref}
\end{figure*}
\begin{figure*}
\centering
\subfloat[Bonfire]{{\includegraphics[width=0.24\textwidth]{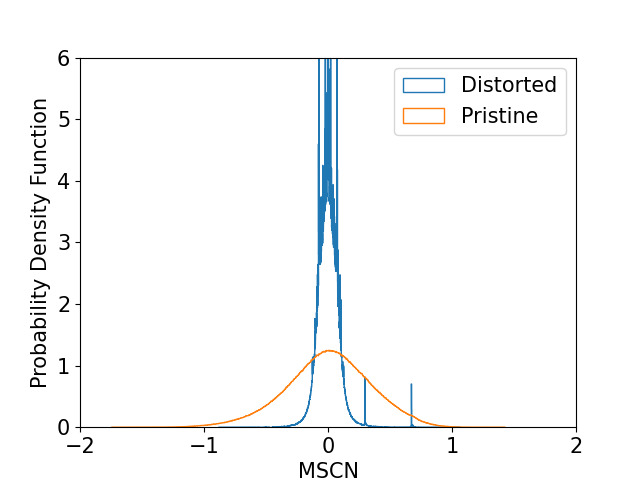}}}
\subfloat[Football]{{\includegraphics[width=0.24\textwidth]{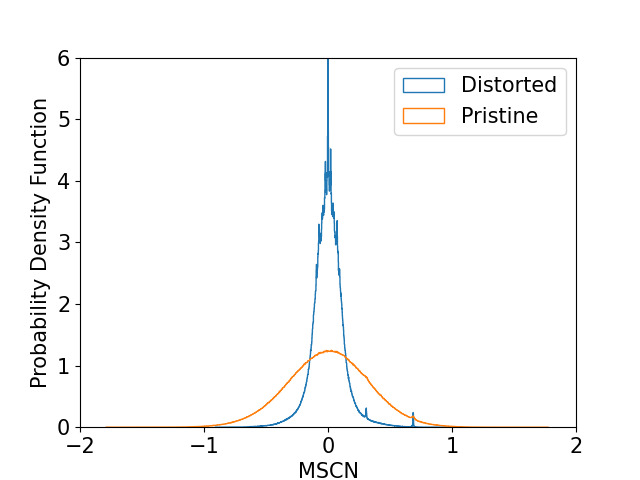}}} 
\subfloat[Golf]{{\includegraphics[width=0.24\textwidth]{{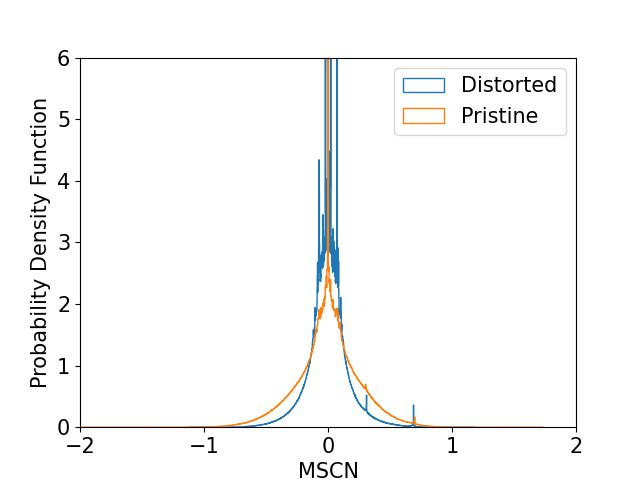}}}}
\subfloat[Fireworks]{{\includegraphics[width=0.24\textwidth]{{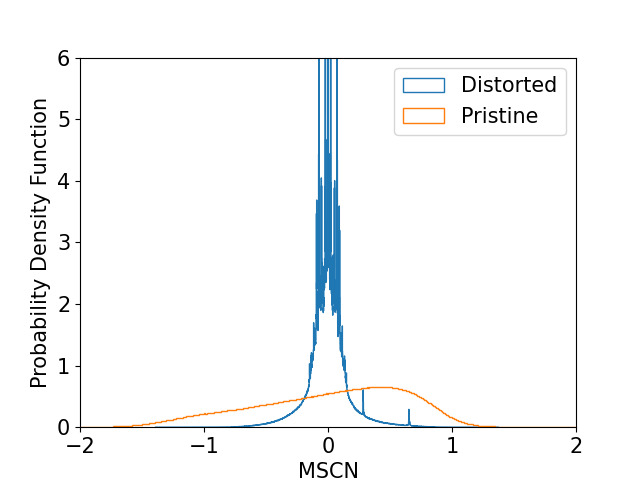}}}} \\
\subfloat[Bonfire after $f(x;4)$]{{\includegraphics[width=0.24\textwidth]{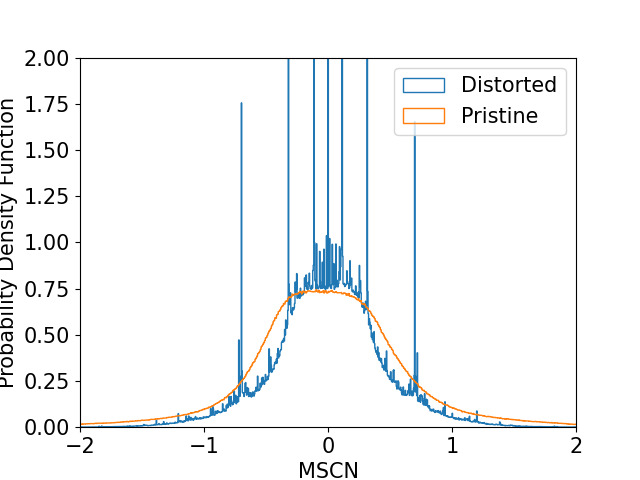}}} 
\subfloat[Football after $f(x;4)$]{{\includegraphics[width=0.24\textwidth]{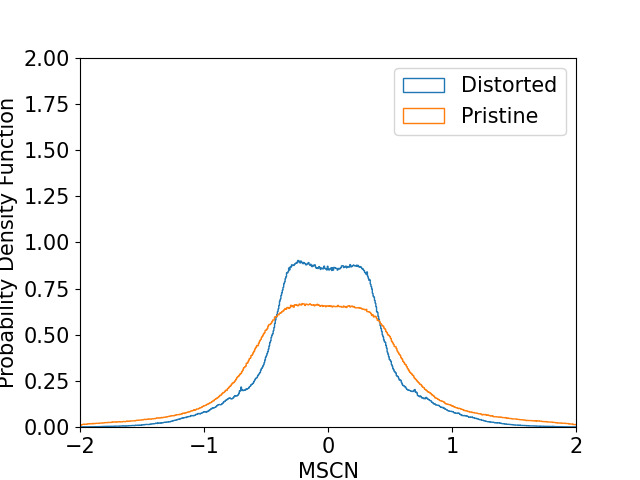}}} 
\subfloat[Golf after $f(x;4)$]{{\includegraphics[width=0.24\textwidth]{{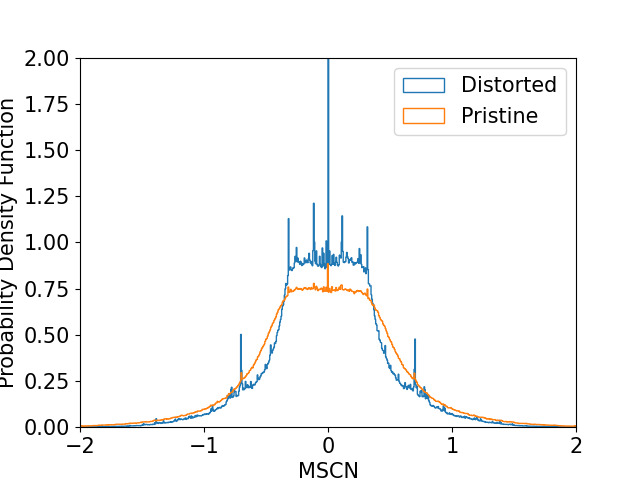}}}}
\subfloat[Fireworks after $f(x;4)$]{{\includegraphics[width=0.24\textwidth]{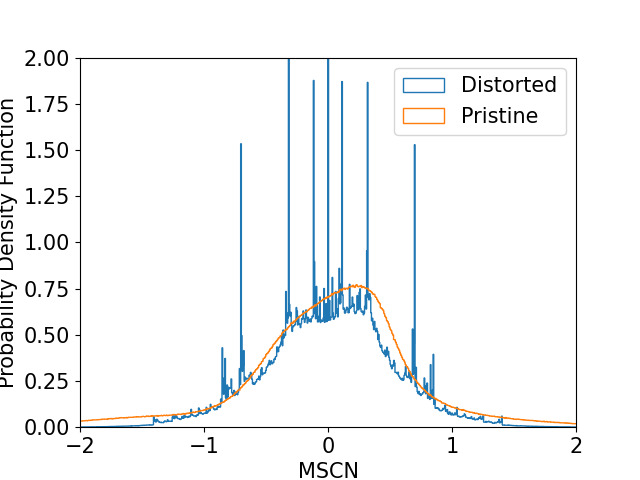}}}
\caption{{Empirical distributions of MSCN coefficients of pristine (blue) and distorted (orange) HDR frames. The first row shows distributions of the MSCN coefficients of luma frames. The second row shows distributions of the MSCN coefficients after $f(x;4)$ ((\ref{eq:exp_piecewise}) with $\delta=4$) was applied to non-overlapping patches of the corresponding frames in the first row.}}
\label{fig:exp_ggd_ref_and_dis}
\end{figure*}

{After (\ref{eq:exp_piecewise}) is applied},  MSCN coefficients are computed on the nonlinearly processed luma and color channels of each processed video frame. We have found that the MSCN coefficients of the nonlinearly processed luma and R$^\prime$G$^\prime$B$^\prime$  values also reliably follow suitably fitted GGD models. The MSCN coefficients are computed using (\ref{eq:mscn}), where we fixed $C=0.001$, since the MSCN coefficients of pixel values processed by (\ref{eq:exp_piecewise}) tend to tightly cluster about the origin. A few plots of the MSCN coefficients of the nonlinearly mapped luma values from pristine frames, with their best GGD fits superimposed on them are shown in Fig.~\ref{fig:exp_ggd_ref}, { while MSCN plots of both pristine and distorted frames from the videos ``Bonfire," ``Football," ``Golf," and ``Fireworks," before and after they were subjected to (\ref{eq:exp_piecewise}), are shown in Fig.~\ref{fig:exp_ggd_ref_and_dis}}. The GGDs fits are apparently quite good, and it may be observed that the statistics of the distorted HDR videos noticeably deviate from those of pristine HDR videos. {As may be seen from Fig.~\ref{fig:exp_ggd_ref_and_dis}, the application of (\ref{eq:exp_piecewise}) causes the MSCN coefficients of the distorted and pristine video frames to become flatter and less concentrated about the mean. The distributions are more spread out, and have larger variances and flatter tops due to the nonlinear operation. The subsequent feature responses therefore capture deviations lying further from the local mean, which tend to be more pronounced on the greater dynamic range of HDR. } 
\par
We have also found that the products of spatially adjacent MSCN coefficients of nonlinearly-processed HDR frames are well described as AGGD.  The shape parameter and variance of the best GGD fits to the MSCN coefficients, and the four parameters of the AGGD fits to the histograms of each of the four products of spatially adjacent MSCN coefficients, are all extracted and used as quality-aware features. As in BRISQUE, all of these features are also computed at a reduced scale by downsampling gaussian-smoothed frames by a factor of 2. This allows the trained prediction model to capture naturally multiscale attributes of videos, and of distorted versions of them.
 
\subsection{Color Component}
~\label{subsection:color}

The BT 2020 color gamut, which is the color gamut that the HDR10 format uses, spans more than twice the range of the SDR color gamut BT 709. While a WCG provides more diverse, brilliant, and realistic colors, it can also increase the visibility of color distortions relative to SDR. For example, color bleeding is an important category of distortions that affect HDR content. It manifests as the smearing of color between areas of strongly contrasting chrominance, with the result that luminance and chrominance edges may not necessarily coincide~\cite{cb1,cb2}. Color bleeding is caused by subsampling of the chromatic channels, and quantization of higher frequency components~\cite{cb3}, and is more prominent in HDR than in SDR content due to the wider range of colors represented in HDR~\cite{cb4}. A WCG also poses challenges to compression and quantization algorithms, because blocking and ringing artifacts can become more visible on brighter colors.



SDR VQA algorithms such as ChipQA~\cite{chipqa} and RAPIQUE \cite{rapique} make use of chrominance components and relative saturation defined using opponent color spaces such as CIELAB. Color spaces that have been designed for HDR content and are intended to be more physiologically plausible than CIELAB include HDR-CIELAB, HDR-IPT~\cite{hdrcielab}, and $J_za_zb_z$~\cite{jzazbz}.  The chrominance components in these opponent color spaces are defined as difference signals whose extremes may not correspond to the extremes of HDR color content. Relative saturation, defined as the magnitudes of color-opponent channels, and referred to as ``chroma" in CIELAB, decouples color from brightness, which is not desirable when studying HDR color artifacts. Amplifying the extremes of chromatic difference signals or relative saturation therefore does not effectively isolate or enhance distortions that arise in areas where the extra bits allocated to HDR produce more extreme brights, colors, and darks that are affected differently by distortions.
\par {Instead, we have found that statistical features extracted from perceptually-uniform R$^\prime$G$^\prime$B$^\prime$  color intensities are remarkably predictive of HDR video quality. Color intensity values captured by camera sensors undergo transformation by the PQ OETF, yielding perceptually uniform R$^\prime$G$^\prime$B$^\prime$ . Rather than applying a nonlinear transformation on color difference signals, we directly analyze the statistics of R$^\prime$G$^\prime$B$^\prime$  values that have been locally mapped on $W\times W$ windows to $[-1,1]$, then subjected to (\ref{eq:exp_piecewise}).}
\par
\begin{figure*}
\centering
\subfloat[Blue channel]{\includegraphics[width=0.24\textwidth]{{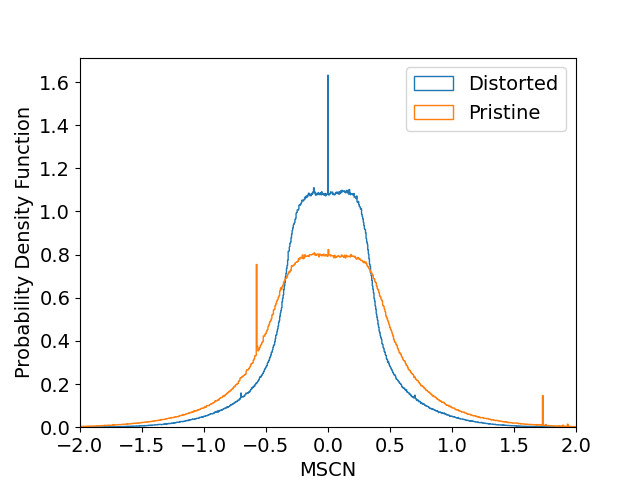}}}
\subfloat[Green channel]{{\includegraphics[width=0.24\textwidth]{jpeg_figs/rgb_mscnNL/540p_2.2M_Flowers_upscaled_rgbnl_mscn_1.jpg}}}
\subfloat[Green channel]{{\includegraphics[width=0.24\textwidth]{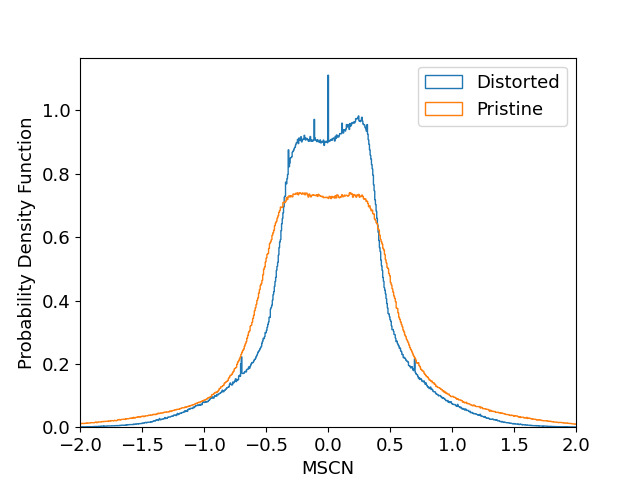}}}
\subfloat[Red channel]{\includegraphics[width=0.24\textwidth]{{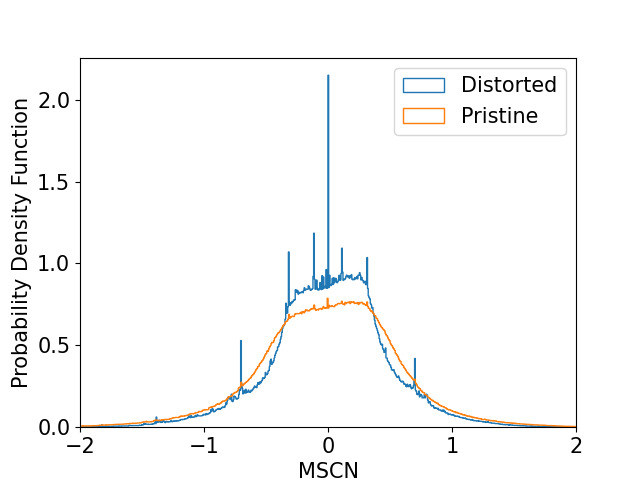}}}
\caption{Empirical distributions of MSCN coefficients of nonlinearly processed R$^\prime$G$^\prime$B$^\prime$  channels of randomly chosen frames of pristine HDR videos (blue) and distorted HDR videos (orange), following application of (\ref{eq:exp_piecewise}) with $\delta=4$. }
\label{fig:rgb_nl_refanddis}
\end{figure*}

As before, assume that the distributions of the MSCN coefficients of the nonlinearly processed R$^\prime$G$^\prime$B$^\prime$  values obey GGD models and after finding the best fits, extract their shape and variance as quality-aware features. Exemplar histograms from different color channel of different randomly selected frames of pristine and distorted HDR content are shown in Fig.~\ref{fig:rgb_nl_refanddis}. We again model  the products of pairs of adjacent MSCN coefficients of each nonlinearly transformed color channel as obeying an AGGD law, and extract the same quality-sensitive features. In order to access ``non-HDR" color distortions, we also process each original R$^\prime$G$^\prime$B$^\prime$  plane, compute their MSCN coefficients, assume their first-order distributions follow GGD models, and that the products of adjacent coefficients follow AGGD models. 

\subsection{Spatio-temporal Components}
\subsubsection{Measuring temporal quality variations}

The `quality-aware' features described in Sections~\ref{subsection:Luma} and~\ref{subsection:color} are extracted on every frame of each video. Specifically, 36 features are extracted on luma, 36 on nonlinearly processed luma (using (\ref{eq:exp_piecewise}) with $\delta=4$), 36 on each R$^\prime$G$^\prime$B$^\prime$  channel, and 36 on each nonlinearly processed R$^\prime$G$^\prime$B$^\prime$  channel, yielding a total of 288 features per frame. We also compute the standard deviation of these features over every non-overlapping five-frame interval, and average them over the video, yielding an additional single descriptor of the variations of quality that may occur.

\subsubsection{ST Gradient Chips}

Visual signals incident on the retina are subjected to a form of spatial bandpass processing and adaptive gain control~\cite{simoncelli}, which are efficiently modelled by the MSCN processing described in (\ref{eq:mscn}). Subsequent stages along the visual pathway perform temporal entropy reduction~\cite{simoncelliMT,adelson,lgn}. The visual signal then passes to area V1, where neurons are sensitive to specific space-time directions. Motivated by the localized spatio-temporal bandpass processing that occurs along the later stages of the retino-cortical pathway, we deploy the ST chip concept to capture the natural statistics of videos being quality-analyzed, along motion-sensitized space-time orientations. 
\par 
Gradients contain important information about edges and amplify distortions that manifest as variations of contrast. In HDR-ChipQA, the gradient magnitudes of the PQ luma frames are first computed using a Sobel operator. Following~\cite{chipqa}, we then apply a discrete temporal filter to the MSCN coefficients of the gradient magnitudes to temporally decorrelate them. 
\par
\begin{figure*}
\centering
\subfloat[]{{\includegraphics[width=0.24\textwidth]{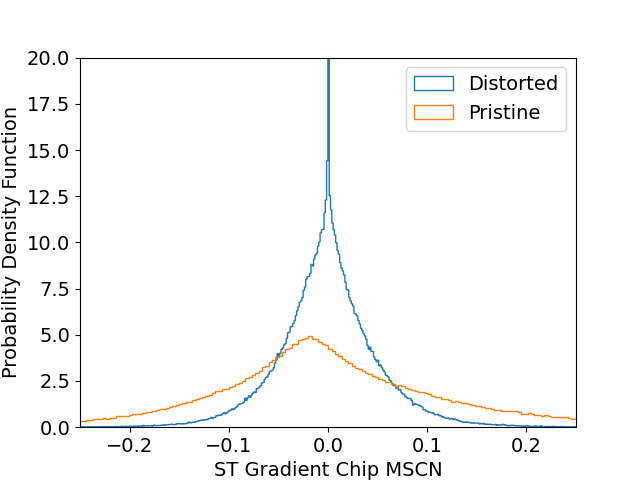}}}
\subfloat[]{{\includegraphics[width=0.24\textwidth]{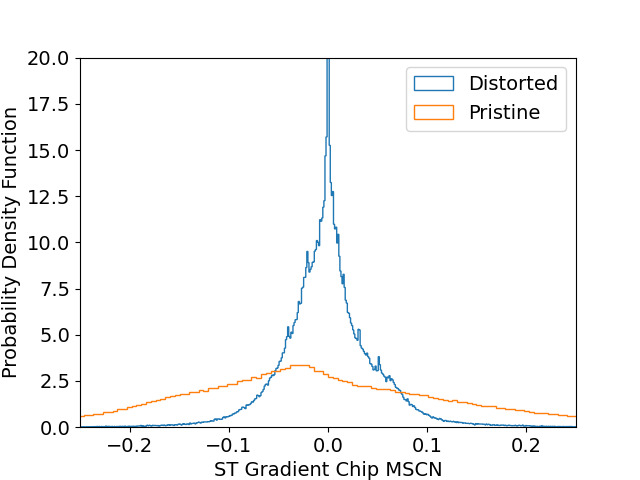}}}
\subfloat[]{{\includegraphics[width=0.24\textwidth]{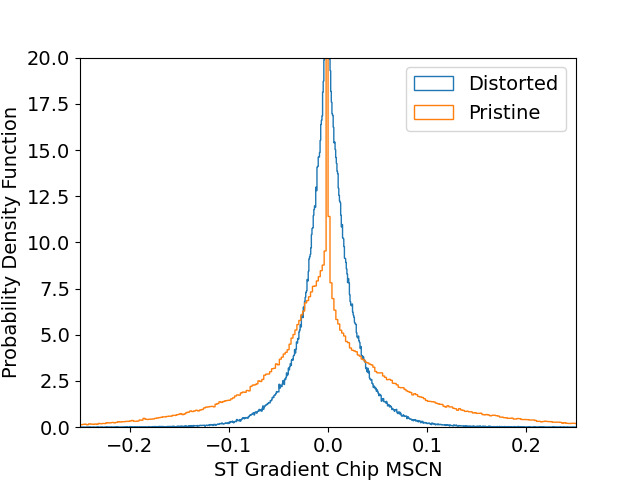}}}
\subfloat[]{{\includegraphics[width=0.24\textwidth]{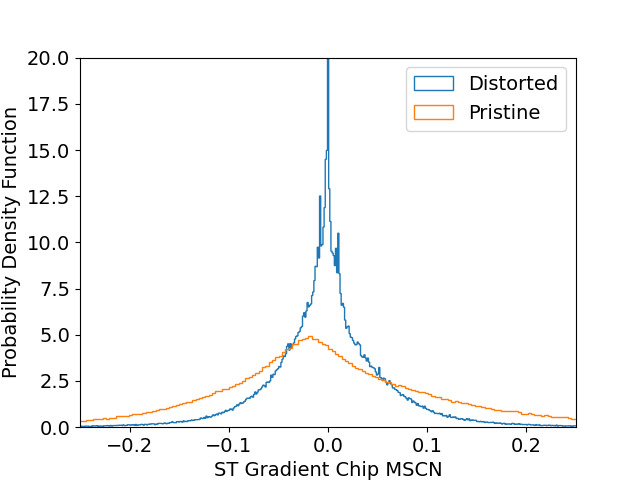}}}
\caption{Empirical distributions of ST Gradient chips of randomly chosen pristine HDR videos (blue) and distorted HDR videos (orange).}
\label{fig:grad_chip_refdis}
\end{figure*}

 
The ST chips of the temporally-processed gradient MSCN coefficients are extracted along the directions of minimum excess kurtosis. The histograms of ST Gradient chips of pristine and distorted HDR videos are plotted together in Fig.~\ref{fig:grad_chip_refdis}. We also computed the paired products of ST Gradient chips and find that they are well modeled as following AGGD probability laws. As before, parameters of the best AGGD fits are used as predictive quality features in the HDR-ChipQA model. For more details, we refer the reader to \cite{chipqa}.

\subsection{Quality Assessment}

A full list and description of the features that define HDR-ChipQA is given in Table~\ref{tab:feats}. We trained an SVR  on all of these features against subjective quality scores, as will be explained.

\begin{table*}
\caption{Descriptions of features in HDR-ChipQA.}
\begin{center}
\resizebox{\textwidth}{!}{
\begin{tabular}{|l|l|l|}
\hline
Domain &  Description & Feature index\\
\hline
Luma &  GGD and AGGD features from HDR Luma. & $f_1-f_{36}$ \\
\hline
Luma after $f(x)$ &  GGD and AGGD features from HDR Luma after nonlinearity  & $f_{37}-f_{72}$  \\
\hline
R$^\prime$G$^\prime$B$^\prime$  & GGD and AGGD features from each channel of R$^\prime$G$^\prime$B$^\prime$ . & $f_{73}-f_{180}$ \\
\hline
R$^\prime$G$^\prime$B$^\prime$   after $f(x)$ & GGD and AGGD features from each channel of R$^\prime$G$^\prime$B$^\prime$  after nonlinearity. & $f_{181}-f_{288}$ \\
\hline
Luma &  Standard deviation every 5 frames of GGD and AGGD features from HDR Luma. & $f_{289}-f_{324}$ \\
\hline
Luma after $f(x)$ &  Standard deviation every 5 frames of GGD and AGGD features from HDR Luma after nonlinearity. & $f_{325}-f_{360}$ \\
\hline
R$^\prime$G$^\prime$B$^\prime$  & Standard deviation every 5 frames of GGD and AGGD features from each channel of R$^\prime$G$^\prime$B$^\prime$ . & $f_{361}-f_{468}$ \\
\hline
R$^\prime$G$^\prime$B$^\prime$   after $f(x)$ & Standard deviation every 5 frames of GGD and AGGD features from each channel of R$^\prime$G$^\prime$B$^\prime$  after nonlinearity. & $f_{469}-f_{576}$ \\
\hline
ST Gradient Chips &  GGD and AGGD features from ST Chips of the gradient magnitude of luma. & $f_{577}-f_{612}$ \\
\hline
\end{tabular}}
\label{feat}
\end{center}
\label{tab:feats}
\end{table*}

\section{Experiments and Results}

\subsection{Performance Metrics}

We computed Spearman's Rank Ordered Correlation Coefficient (SROCC) between the scores predicted by HDR-ChipQA and ground truth mean opinion scores (MOS). We also fit the predicted scores to the MOS using a logistic function
\begin{equation}
l(s) =  \frac{\beta_1-\beta_2}{ 1 + \exp(-\frac{(x - \beta_3)}{ \beta_4}) + \beta_5} 
\end{equation}
and then computed Pearson's Linear Correlation Coefficient (LCC) and the Root Mean Square Error (RMSE) between the fitted scores and the MOS, following standard practice~\cite{logistic}.

\subsection{Databases}
{To the best of our knowledge, there currently exists only one publicly-available HDR10 database: the LIVE-APV HDR~(\cite{livehdr}) database. The DML-HDR database~\cite{dmlhdr} contains 10 bit videos, but they are expressed in the BT709 color space and are hence not compliant with modern HDR standards such as HDR10, HDR10+, or Dolby Vision. In addition, the scores for this database are not publicly available. The HDR-HDM-2014 database~\cite{stuttgart} has been used to evaluate tone-mapping operators but no subjective study has been conducted using this dataset and thus it has no quality scores. } We evaluated HDR-ChipQA against other models on the LIVE-APV HDR database. LIVE-APV HDR contains 310 videos that were viewed by human subjects under two ambient conditions. The videos were created by applying {9 different combinations of compression and downsampling } on 31 source videos. The ambient conditions were a dark setting with an incident luminance of $<10$ lux, and a bright setting with an incident lumination of 200 lux. The scores that were recorded under the two ambient settings were found to be statistically indistinguishable from each other. We separately evaluated HDR-ChipQA on both sets of scores. The scores were retrieved from raw opinion scores using the Sureal method of maximum likelilood estimation~\cite{sureal}.

{In order to evaluate its robustness against dynamic range, we also evaluated it on the 10-bit SDR LIVE ETRI~\cite{etri} database and the 8-bit SDR LIVE Livestream~\cite{livestream} database, and these results are reported in Section IV.G.}

\subsection{Evaluation Protocol}
We trained a Support Vector Regressor (SVR) with a linear kernel on the features to predict the MOS of the videos. The training process was as follows: We partitioned the database into a training set and a test set with an 80:20 ratio so that all videos of a same content appear in the same set. This content separation was done to prevent the SVR from learning any content-specific cues from the features, which could artificially inflate performance. We performed 5-fold cross validation on the training set to choose the best value of the regularization parameter $C$ of the SVR. No hyperparameter tuning was done on the test set. This procedure was repeated 100 times and the median metrics are reported for each database.
\subsection{Results on LIVE HDR Database}

 We evaluated HIGRADE, TLVQM, V-BLIINDS, HDR-BVQM, {VSFA}, RAPIQUE, (original) ChipQA, and BRISQUE. All of these models were trained on the LIVE-HDR database for a fair comparison. {We also attempted to train the NorVDPNet network on the LIVEHDR database but it failed to converge. We hence evaluated NorVDPNet using weights obtained by pretraining the network on HDR images having JPEG-XT distortions~\cite{jpeg}. Note that VSFA is also a CNN-based metric, and RAPIQUE and VIDEVAL are hybrids of CNN models with feature-based models}.  HIGRADE, TLVQM, and ChipQA were originally defined on features computed from the CIELAB colorspace. For fairer comparison, we instead trained all models using the HDR-CIELAB colorspace.  The performance results are presented in Table~\ref{tab:results} for scores gathered under the dark viewing condition, and in Table~\ref{tab:bright_results} for scores gathered under the bright viewing condition. HDR-ChipQA outperformed all of the other algorithms by a wide margin, obtaining a nearly 10\% increase over the next-best performing algorithm. 

\begin{table*}[!h]
\caption{Median SROCC, LCC, and RMSE on the LIVE HDR Database on scores collected under the dark ambient condition for NR algorithms. Standard deviations are shown in parentheses. The best performing algorithm is bold-faced.}
\begin{center}
\begin{tabular}{|c|c|c|c|c|}
\hline
\multicolumn{2}{|c|}{\textsc{Method}}  &  SROCC$\uparrow$ & LCC$\uparrow$ & RMSE$\downarrow$  \\  
\hline
\multirow{3}{*}{\textsc{Image Quality Metrics}} 
& RAPIQUE \cite{rapique} & 0.4553 \scriptsize(0.2533) & 0.4864 \scriptsize(0.1171) & 15.7134 \scriptsize(1.7415) \\
\cline{2-5}
& {NoR-VDPNet}\cite{norvdpnet} & 0.6248\scriptsize(0.0635) & 0.5605\scriptsize(0.0764) & 14.8226 \scriptsize(2.1892) \\
\cline{2-5}
& HIGRADE \cite{higrade} & 0.7088 \scriptsize(0.0827) & 0.6827 \scriptsize(0.0710) & 14.2545 \scriptsize(2.0780) \\
\cline{2-5}
& BRISQUE\cite{brisque} &  0.7251 \scriptsize(0.0955) & 0.7139 \scriptsize(0.0881) & 12.6404  \scriptsize(2.1651)  \\
\hline
\multirow{6}{*}{\textsc{Video Quality Metrics}}
& TLVQM \cite{tlvqm} &  0.5781 \scriptsize(0.1014) & 0.5552 \scriptsize(0.0919) & 14.999 \scriptsize(1.9098)\\
\cline{2-5}
& HDR BVQM\cite{hdrbvqm} & 0.6020 \scriptsize(0.0944) & 0.5844 \scriptsize(0.086) & 14.5930 \scriptsize(1.8276) \\
\cline{2-5}
& {VSFA}\cite{vsfa} & 0.7127 \scriptsize(0.1079) & 0.6918 \scriptsize(0.1114) & 13.0511 \scriptsize(2.4003) \\
\cline{2-5}
& V-BLIINDS\cite{vbliinds} & 0.7483 \scriptsize(0.1446) & 0.7193 \scriptsize(0.1141) & 12.7794 \scriptsize(2.3715) \\
\cline{2-5}
& ChipQA~\cite{chipqa} & 0.7435 \scriptsize(0.0895) & 0.7334 \scriptsize(0.0819) & 12.1549 \scriptsize(1.9106) \\
\cline{2-5}
& \textbf{HDR-ChipQA} &  \textbf{0.8250\scriptsize(0.0589)} & \textbf{0.8344\scriptsize(0.0562)} & \textbf{9.8038 \scriptsize(1.7334)} \\
\hline
\end{tabular}
\label{tab:results}
\end{center}
\vspace{-5mm}
\end{table*}

\begin{table*}[!h]
\caption{Median SROCC, LCC, and RMSE on the LIVE HDR Database on scores collected under the bright ambient condition for NR algorithms. Standard deviations are shown in parentheses. The best performing algorithm is bold-faced.}
\begin{center}
\begin{tabular}{|c|c|c|c|c|}
\hline
\multicolumn{2}{|c|}{\textsc{Method}}  &  SROCC$\uparrow$ & LCC$\uparrow$ & RMSE$\downarrow$  \\  
\hline
\multirow{4}{*}{\textsc{Image Quality Metrics}} 
& RAPIQUE \cite{rapique} & 0.4470 \scriptsize(0.2171) &  0.4910\scriptsize(0.1393) &  15.6088\scriptsize(1.9382) \\
\cline{2-5}
& {NoR-VDPNet}\cite{norvdpnet} & 0.5753 \scriptsize(0.0599) & 0.5383 \scriptsize(0.0616) & 14.5622 \scriptsize(1.8899) \\
\cline{2-5}
& HIGRADE \cite{higrade} & 0.6862 \scriptsize(0.0973) & 0.6664 \scriptsize(0.0808) & 13.7339\scriptsize(2.0078)  \\
\cline{2-5}
& BRISQUE\cite{brisque} &  0.7133 \scriptsize(0.1004) &  0.7139\scriptsize(0.0885) & 12.6404\scriptsize(2.0428)  \\
\hline
\multirow{5}{*}{\textsc{Video Quality Metrics}}
& HDR BVQM\cite{hdrbvqm} & 0.5411 \scriptsize(0.1102) & 0.5436 \scriptsize(0.0986) & 15.4146 \scriptsize(1.8312)  \\
\cline{2-5}
& TLVQM \cite{tlvqm} &  0.5549 \scriptsize(0.1162) & 0.5504 \scriptsize(0.1008) & 15.2480 \scriptsize(1.8562)\\
\cline{2-5}
& V-BLIINDS\cite{vbliinds} & 0.7248 \scriptsize(0.1304) & 0.7009 \scriptsize(0.1180) & 12.896 \scriptsize(2.3606) \\
\cline{2-5}
& ChipQA~\cite{chipqa} & 0.7437 \scriptsize(0.0815) &0.7312 \scriptsize(0.0864) & 12.3509 \scriptsize(1.843) \\
\cline{2-5}
& \textbf{HDR-ChipQA} &  \textbf{0.8316\scriptsize(0.0580)} & \textbf{0.8287 \scriptsize(0.0552)} & \textbf{10.1903 \scriptsize(1.6664)} \\
\hline
\end{tabular}
\label{tab:bright_results}
\end{center}
\end{table*}

\begin{figure*}[h]
\centering
\subfloat[RAPIQUE]{{\includegraphics[width=0.22\textwidth]{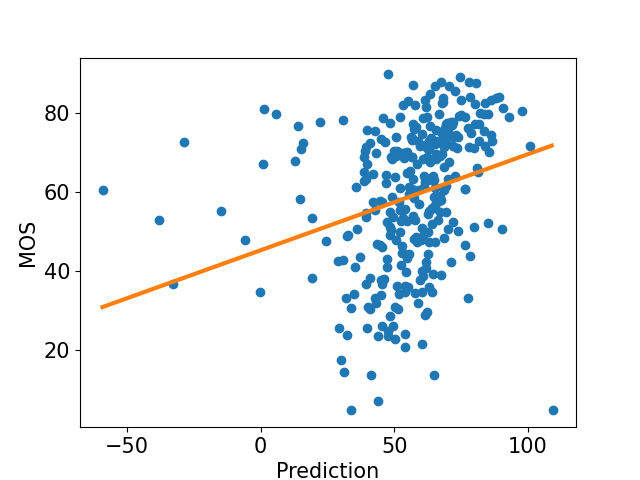}}}
\subfloat[TLVQM]{{\includegraphics[width=0.22\textwidth]{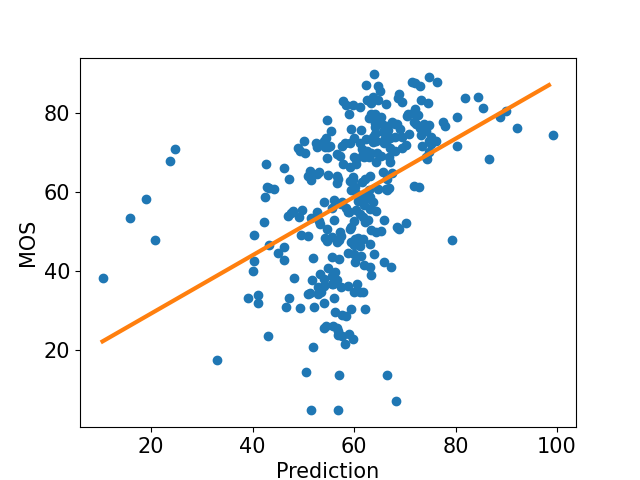}}}
\subfloat[HDR BVQM]{{\includegraphics[width=0.22\textwidth]{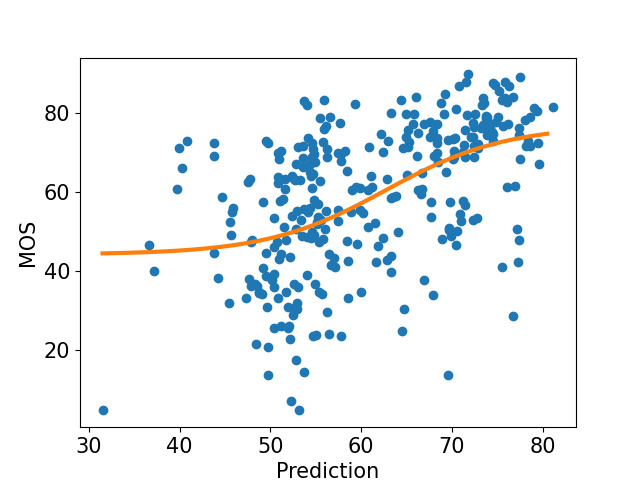}}}
\subfloat[NoR-VDPNet]{{\includegraphics[width=0.2\textwidth]{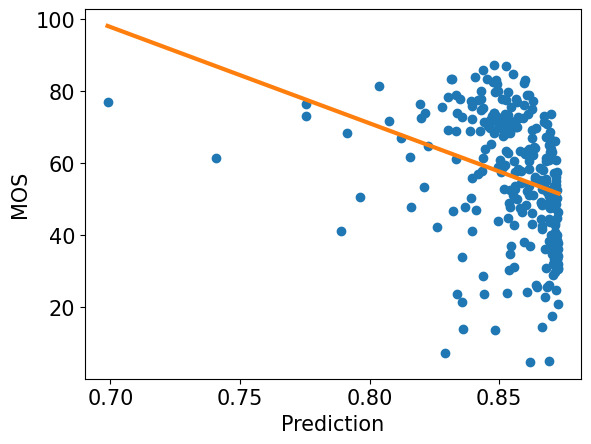}}}\\
\subfloat[HIGRADE]{{\includegraphics[width=0.22\textwidth]{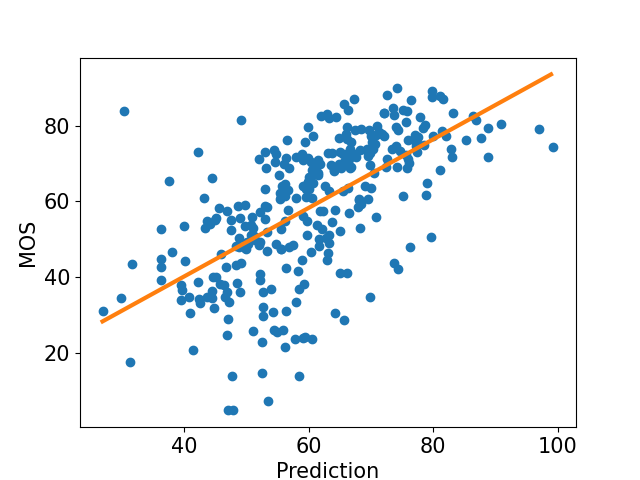}}}
\subfloat[VSFA]{{\includegraphics[width=0.2\textwidth]{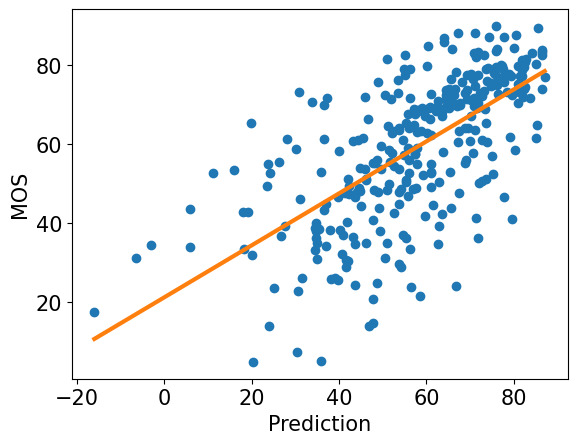}}}
\subfloat[BRISQUE]{{\includegraphics[width=0.22\textwidth]{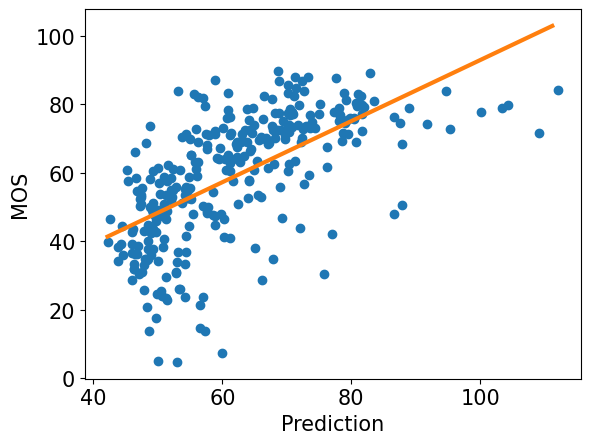}}}
\subfloat[VBLIINDS]{{\includegraphics[width=0.22\textwidth]{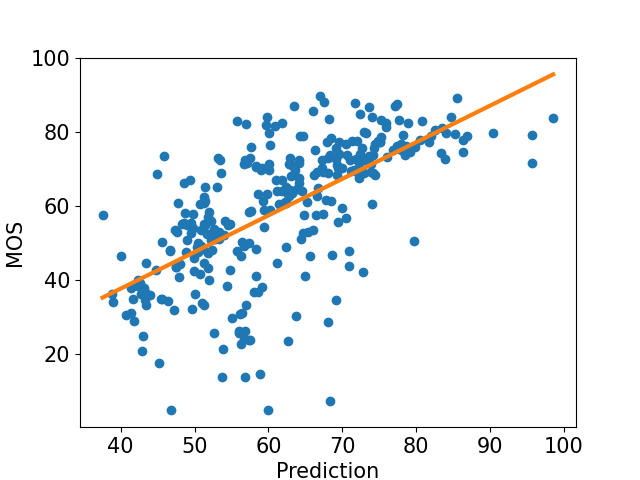}}}\\
\subfloat[ChipQA]{{\includegraphics[width=0.24\textwidth]{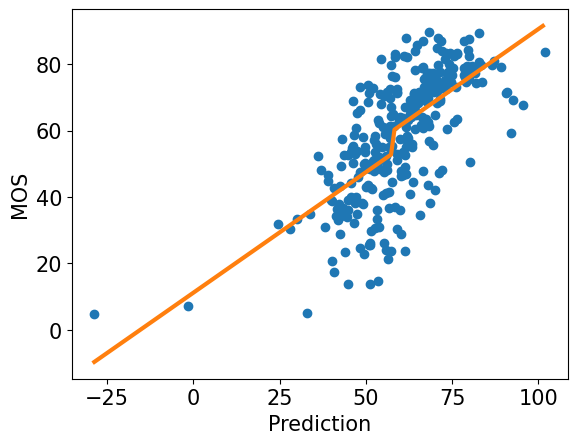}}}
\subfloat[HDR-ChipQA]{{\includegraphics[width=0.24\textwidth]{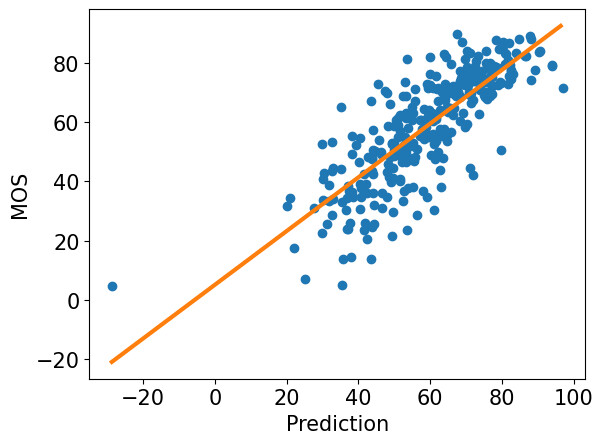}}}
\caption{Scatter Plots of  MOS vs model predictions, with parametric fits $l(s)$ shown in orange.}
\label{fig:scatter}
\end{figure*}

\begin{figure*}[h]
\centering
\includegraphics[width=0.8\textwidth]{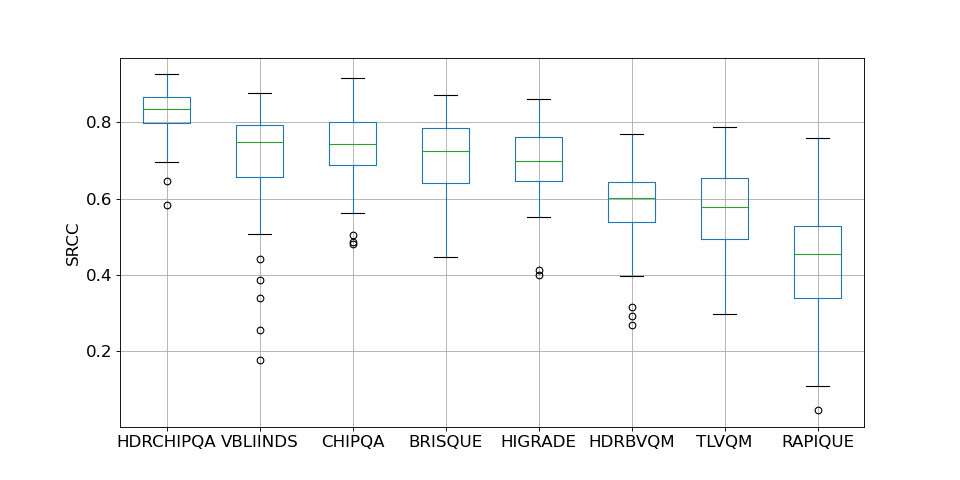}
\caption{Boxplot of SRCC values attained by the compared models over 100 splits}
\label{fig:boxplot}
\end{figure*}

Scatter plots of the predictions made by all of the compared algorithms on the videos obtained under the dark viewing condition are shown in Fig.~\ref{fig:scatter}. This visualization clearly shows that HDR-ChipQA makes predictions that are more consistent with MOS than those produced by the other models. The scatter plots were created by plotting the MOS against the mean quality predictions produced by the NR VQA algorithms on each video in the test set over the 100 train-test splits, ensuring that each video appeared in the test split at least once. A boxplot of the SRCC values obtained over the 100 train-test splits is shown in Fig.~\ref{fig:boxplot}, showing that HDR-ChipQA attained a higher median SRCC and a smaller spread of SRCC than the other algorithms.


 V-BLIINDS, BRISQUE, and ChipQA achieved similar performances on the LIVE HDR database. TLVQM is a state-of-the-art algorithm but performs very poorly on HDR. This might be due to the large number of parameters in TLVQM that are defined and tested by the authors of \cite{tlvqm} on SDR content. RAPIQUE is a state-of-the-art SDR VQA algorithm but it also performed poorly on LIVE HDR. HDR-ChipQA, on the other hand, was able to achieve high performance against all performance metrics.
 \par
 We also performed a one-sided Welch's t-test with a 95\% confidence interval on the SRCC values obtained by evaluating various algorithms on the scores collected under the dark viewing condition {over 100 random train-test splits. We chose the Welch's t-test over the Student's t-test because the Welch's t-test does not assume that the variances of the distributions of SRCC values for each algorithm are equal, which Student's t-test assumes and is incorrect in this case~\cite{welch}.} The results are shown in Table~\ref{tab:ttest}. Although some of the algorithms attained SRCC values that were statistically indistinguishable from each other, the results again show the elevated performance and statistical superiority of HDR-ChipQA against all the other algorithms tested.

\begin{table*}[h]
\caption{Results of one-sided t-test performed between SRCC values of the compared algorithms on the LIVE-HDR database. `1' (`-1') indicates that the row algorithm is statistically superior (inferior) to the column algorithm. 0 indicates that the null hypothesis that the mean SRCC was the same for the row and column algorithm cannot be rejected. The matrix is symmetric.}
\begin{center}
\resizebox{\textwidth}{!}{
\begin{tabular}{|l|l|l|l|l|l|l|l|l|l|l|}
\hline
\textsc{Method}  & RAPIQUE & HDR BVQM & TLVQM & {NoR-VDPNet} &  {VSFA} &  HIGRADE & BRISQUE & V-BLIINDS & ChipQA & HDR-ChipQA   \\
\hline
RAPIQUE & - &  -1 & -1 & -1 & -1 & -1 & -1 & -1 & -1 & -1  \\
\hline
HDR BVQM & 1 & - & 0 & 0 & -1 & -1 & -1 & -1 & -1 & -1  \\
\hline
TLVQM & 1 & 0 & - & 0 & -1 & -1 & -1 & -1 & -1  & -1  \\
\hline 
{NoR-VDPNet}  &  1 & 0 & 0 & - & -1 & -1 & -1 & -1 & -1  & -1   \\
\hline
{VSFA} & 1 & 1 & 1 & 1 & - & 0 & 0 & 0  & -1  & -1  \\
\hline 
HIGRADE  &  1 & 1 & 1 & 1 & 0 & - & 0 & 0 & -1  & -1   \\
\hline
BRISQUE & 1 & 1 & 1 & 1 & 0 & 0 & - & 0 & -1 & -1  \\
\hline
V-BLIINDS &  1 & 1 & 1 & 1 & 0 & 0 & 0 & - & 0 & -1   \\
\hline
ChipQA &  1 & 1 & 1 & 1 & 1 & 1 & 1 & 0 & - & -1  \\
\hline
HDR-ChipQA &  1 & 1 & 1 & 1 & 1 & 1 & 1 & 1 & 1 &  -  \\
\hline
\end{tabular}}
\label{tab:ttest}
\end{center}
\end{table*}

 \subsection{Ablation Study}

 We also conducted a systematic ablation study by removing each feature space from the algorithm, then evaluating the performance of the rest of the algorithm on the scores collected under the dark viewing condition. The results are presented in Table~\ref{tab:ablation}, with the feature spaces ranked by the drop of SRCC they caused when they were removed. 
 
 \begin{table}
\caption{Results of ablation study. Median SROCC over 100 splits on the LIVE HDR VQA database, as different feature spaces were removed}
\begin{center} 
\begin{tabular}{|c|c|c|}
\hline
\textsc{Rank} & \textsc{Feature Spaces Removed}  & SROCC \\
\hline
1 & Statistical features of nonlinear luma & 0.7691 \\
\hline
2 &  Statistical features of ST Gradient Chips & 0.8151 \\
\hline
3 & Statistical features of nonlinear R$^\prime$G$^\prime$B$^\prime$  & 0.8257 \\
\hline
4 & Statistical features of luma &  0.8262 \\
\hline
5 & Statistical features of R$^\prime$G$^\prime$B$^\prime$  &  0.8357 \\
\hline
\end{tabular}
\label{tab:ablation}
\end{center}
\vspace{-5mm}
\end{table}

We also evaluated the performance of each feature space separately and reported the median SRCC when each feature space was trained and tested to predict human scores gathered under the dark viewing condition in Table~\ref{tab:separate}. These experiments indicate that the features drawn from the nonlinearly processed luma and R$^\prime$G$^\prime$B$^\prime$  spaces were understandably the most important for HDR quality assessment.
 \begin{table}[!h]
\caption{Median SROCC for 100 splits on the LIVE HDR VQA database for each feature space}
\begin{center} 
\begin{tabular}{|c|c|c|}
\hline
\textsc{Rank} & \textsc{Feature Space}  & SROCC \\
\hline
1 &  Statistical features of nonlinear luma  & 0.8141 \\ 
\hline
2 & Statistical features of nonlinear R$^\prime$G$^\prime$B$^\prime$  & 0.7470   \\ 
\hline
3 & Statistical features of R$^\prime$G$^\prime$B$^\prime$    & 0.7273 \\
\hline
4 &   Statistical features of luma & 0.7016  \\ 
\hline
5 & ST Gradient Chips  & 0.4599\\
\hline
\end{tabular}
\label{tab:separate}
\end{center}
\vspace{-5mm}
\end{table}

\subsection{Experiments on the Expansive Nonlinearity}

We conducted experiments on the patch size $W$ for $\delta=4$. Results of comparing the predictions to the scores collected under the dark viewing condition for $W=9,17,31$ and $63$ are reported in Table~\ref{tab:patch}  {The results were computed on the entire LIVE HDR dataset following the procedure described in section IV C}. We intentionally chose values of $W$ offset by a single pixel from multiples of 4 so they would not align with compression block boundaries. We found that very small patch sizes led to extreme values rendering the mapping to $[-1,1]$ almost meaningless and preventing the enhancement of contrast. We also tried applying the nonlinearity to entire frames instead of locally, obtaining the results labeled as ``Global" in Table X. We found that $W=17$ produced the best results, and that local mapping to $[-1,1]$ before application of (\ref{eq:exp_piecewise}) was more beneficial than global rescaling.

\begin{table}
\caption{Median SROCC, LCC, and RMSE of HDR-ChipQA for 100 splits on the LIVE HDR VQA database for different values of patch size parameter $W$. Standard deviation in parentheses. Best perfoming $W$ is boldfaced.} 
\begin{center} 
\begin{tabular}{|l|l|l|l|l|}
\hline
$W$  & SROCC & LCC & RMSE \\
\hline
9 &  0.7885\scriptsize(0.0829) & 0.7412 \scriptsize(0.0896) & 11.9931 \scriptsize(2.0985)\\
\hline
\textbf{17} &   \textbf{0.8250\scriptsize(0.0589)} & \textbf{0.8344\scriptsize(0.0562)} & \textbf{9.8038 \scriptsize(1.7334)} \\
\hline
31 & 0.7725 \scriptsize(0.0892) & 0.7469 \scriptsize(0.0876) & 11.9691 \scriptsize(2.0698) \\
\hline
63 &  0.7388\scriptsize(0.0898) & 0.6977\scriptsize(0.0885) & 13.1210 \scriptsize(2.1280) \\
\hline
Global & 0.7165\scriptsize(0.1162) & 0.6552\scriptsize(0.1433) & 12.8888 \scriptsize(2.5517) \\ 
\hline
\end{tabular}
\label{tab:patch}
\end{center}
\end{table}

 We also conducted experiments to find the best value of $\delta$ in the nonlinearity $f(x)$ in (\ref{eq:exp_piecewise}) for $W=17$. As discussed in section~\ref{brisque_fx}, we found that larger values of $\delta$ caused the mid-range of luma values to be flattened and the ends of the luma range to be contrast-enhanced, but very large values of $\delta$ caused the flattening of mid-ranges of luma that were still relevant to HDR, while emphasizing luma extremes to a greater degree than necessary. We computed median SRCC values {on the entire dataset following the procedure described in Section IV C} for HDR-ChipQA using $\delta=1,2,3,4,5$ and report the results in Table~\ref{tab:delta}.  . As may be seen, choosing $\delta=4$ delivered the best performance.

 It is to be noted that since these results are the median and standard deviations of metrics computed for 100 splits, they are test-set agnostic.
 
\begin{table}[!h]
\caption{Median SROCC, LCC, and RMSE of HDR-ChipQA over 100 train-test splits on the LIVE HDR VQA database for different values of nonlinearity parameter $\delta$. Standard deviations are given in parentheses. The best perfoming value of $\delta$ is boldfaced.} 
\begin{center} 
\begin{tabular}{|l|l|l|l|l|}
\hline
$\delta$  & SROCC & LCC & RMSE \\
\hline
1 &  0.7464\scriptsize(0.0906) & 0.7097 \scriptsize(0.1040) & 11.3032 \scriptsize(2.2478)\\
\hline
2 & 0.7464\scriptsize(0.0949) & 0.7153 \scriptsize(0.1112) & 13.0305 \scriptsize(2.3490) \\
\hline
3 &  0.7410\scriptsize(0.09391) & 0.7048 \scriptsize(0.0986) & 12.8876 \scriptsize(2.2272) \\
\hline
\textbf{4} &   \textbf{0.8250\scriptsize(0.0589)} & \textbf{0.8344\scriptsize(0.0562)} & \textbf{9.8038 \scriptsize(1.7334)} \\
\hline
5 & 0.8069\scriptsize(0.0906) & 0.7698 \scriptsize(0.1080) & 11.3032 \scriptsize(2.5455)\\
\hline
\end{tabular}
\label{tab:delta}
\end{center}
\end{table}

\subsection{Results on SDR Databases}
 
 We also evaluated HDR-ChipQA on the LIVE Livestream~\cite{livestream} and on the LIVE ETRI databases~\cite{etri}. We evaluated HDR-ChipQA on SDR content by implementing R$^\prime$G$^\prime$B$^\prime$  features using the BT 709 gamut space, instead of the BT 2020 gamut, and implementing luma features using BT.709 gamma-corrected luma instead of PQ luma. {As shown in Tables~\ref{tab:liveapv} and \ref{tab:etri}, we found that HDR-ChipQA achieved the best video quality prediction performances on both the LIVE Livestream dataset and the LIVE ETRI database, respectively.} The state-of-the-art performance of HDR ChipQA on both HDR and SDR databases suggests that it may be used agnostically with respect to dynamic range.  {The strong performance of the algorithm may be due to the fact that image contrast is strongly affected by distortions regardless of dynamic range, and therefore enhancing the local contrast of frames increases the predictive power of subsequent features.}
 \par
{The LIVE HDR dataset contains compression and downsampling distortions. The LIVE Livestream dataset contains blur, noise, flicker, judder, interlacing, compression, and downsampling. The LIVE ETRI dataset contains videos with compression, downsampling, and temporal subsampling. Our proposed model achieves the best NR VQA performance on all of these datasets, showing that it can predict the quality of videos affected by any of these distortions as long as it is suitably trained. }
 \begin{table*}[!h]
\caption{Median SROCC, LCC, and RMSE obtained by the compared models on the LIVE Livestream Database. Standard deviations are shown in parentheses. The best performing algorithm is bold-faced.}
\begin{center}
\begin{tabular}{|c|c|c|c|c|}
\hline
\multicolumn{2}{|c|}{\textsc{Method}}  &  SROCC$\uparrow$ & LCC$\uparrow$ & RMSE$\downarrow$  \\  
\hline
\multirow{6}{*}{\textsc{Image Quality Metrics}} & NIQE\cite{niqe} & 0.3104 \scriptsize(0.0963) & 0.48674 \scriptsize(0.3323) & 11.2799 \scriptsize(1.0362) \\
\cline{2-5}
& VGG-19\cite{vgg19} & 0.5887 \scriptsize(0.0853) &  0.6600 \scriptsize(0.0687) &  9.3951 \scriptsize(0.6587) \\ 
\cline{2-5}
& ResNet-50\cite{resnet} & 0.6395 \scriptsize(0.0793) & 0.7011 \scriptsize(0.0666) & 8.9941 \scriptsize(0.6361) \\
\cline{2-5}
& BRISQUE\cite{brisque} &  0.6564\scriptsize(0.1140) & 0.6840\scriptsize(0.1019) & 9.2822\scriptsize(1.5088)  \\
\cline{2-5}
& HIGRADE\cite{higrade} & 0.7088 \scriptsize(0.0805) & 0.7247 \scriptsize(0.0742) & 9.0958 \scriptsize(1.1249) \\
\cline{2-5}
& CORNIA\cite{cornia}  & 0.6482 \scriptsize(0.0977) & 0.6826 \scriptsize(0.0819) & 9.6472 \scriptsize(1.0725) \\
\hline
\multirow{6}{*}{\textsc{Video Quality Metrics}} &  VIIDEO\cite{viideo} & 0.0005 \scriptsize(0.0965) & 0.2643 \scriptsize(0.1670) & 12.0508 \scriptsize(0.9087) \\ 
\cline{2-5}
& V-BLIINDS\cite{vbliinds} & 0.7464\scriptsize(0.0833) & 0.7572\scriptsize(0.0748) & 8.5551\scriptsize(1.1485) \\
\cline{2-5}
& TLVQM\cite{tlvqm} &  0.7201\scriptsize(0.0993) & 0.7191\scriptsize(0.0965) & 9.3182\scriptsize(1.4947)
)\\
\cline{2-5}
& ChipQA-0\cite{chipqa0} & 0.7852\scriptsize(0.0774) & 0.7951\scriptsize(0.0671) & 7.8975\scriptsize(1.2838) \\
\cline{2-5}
& ChipQA~\cite{chipqa} & 0.7909\scriptsize(0.0764) & 0.8009\scriptsize(0.0694) & 7.8745\scriptsize(1.3425)\\
\cline{2-5}
& \textbf{HDR-ChipQA} &  \textbf{0.8397\scriptsize(0.0635)} & \textbf{0.8570\scriptsize(0.0516)} & \textbf{6.8678\scriptsize(1.2642)} \\
\hline
\end{tabular}
\label{tab:liveapv}
\end{center}
\vspace{-5mm}
\end{table*}

\begin{table*}
\caption{Median SROCC, LCC, and RMSE obtained by the compared models on the LIVE ETRI database. Standard deviations are in parentheses. The best performing algorithm is bold-faced.}
\begin{center}
\begin{tabular}{|c|c|c|c|c|}
\hline
\multicolumn{2}{|c|}{\textsc{Method}}  &  SROCC$\uparrow$ & LCC$\uparrow$ & RMSE$\downarrow$  \\  
\hline
\multirow{2}{*}{\textsc{Image Quality Metrics}} & NIQE\cite{niqe} & 0.3966 \scriptsize(0.1983) & 0.4435 \scriptsize(0.3049) & 12.2663 \scriptsize(1.6530) \\
\cline{2-5}
& BRISQUE\cite{brisque} &  0.2656 \scriptsize(0.3042) & 0.4315 \scriptsize(0.2461) & 12.3339 \scriptsize(1.5116)   \\
\hline
\multirow{5}{*}{\textsc{Video Quality Metrics}} & TLVQM\cite{tlvqm} &  0.2343 \scriptsize(0.1934) & 0.3018 \scriptsize(0.1957) & 13.0663 \scriptsize(1.2398)\\
\cline{2-5}
& V-BLIINDS\cite{vbliinds} & 0.4798 \scriptsize(0.1411) & 0.5344 \scriptsize(0.1194) & 11.7581 \scriptsize(1.4058) \\
\cline{2-5}
& ChipQA-0\cite{chipqa0} & 0.4012 \scriptsize(0.1591) & 0.4634 \scriptsize(0.1356) & 12.1476 \scriptsize(1.3464) \\
\cline{2-5}
& ChipQA~\cite{chipqa} & 0.6323 \scriptsize(0.1474) & 0.6822 \scriptsize(0.1182) & 10.0769 \scriptsize(1.4701) \\
\cline{2-5}
& \textbf{HDR ChipQA} & \textbf{0.7297}\scriptsize(\textbf{0.1258}) & \textbf{0.7275}\scriptsize(\textbf{0.1144}) & \textbf{9.2952}\scriptsize(\textbf{1.7091})\\
\hline
\end{tabular}
\label{tab:etri}
\end{center}
\end{table*}

\section{Conclusion}
We have created the first NR VQA algorithm for HDR10 content. The algorithm is based on principles from natural scene statistics adapted for HDR. We found that an existing NR VQA algorithm designed for SDR content can be improved upon by making use of perceptually-motivated features specifically created for HDR. HDR-ChipQA is able to achieve very high correlations against human judgments of both HDR and SDR video quality, and we envision that it will be a useful tool for a wide variety of applications. {The source code for HDR ChipQA will be made available at https://github.com/JoshuaEbenezer}.

\section*{Acknowledgment}

This research was sponsored by a grant from Amazon.com, Inc., and by grant number 2019844 for the National Science Foundation AI Institute for Foundations of Machine Learning (IFML). The authors also thank the Texas Advanced Computing Center (TACC) at The University of Texas at Austin for providing HPC resources that have contributed to the research results reported in this paper. URL: http://www.tacc.utexas.edu.

\bibliographystyle{model1-num-names}

\bibliography{hdrchipqa}


\end{document}